\documentclass[11pt]{article}

\usepackage{epsf}
\usepackage{color}
\usepackage{amsmath,amssymb,color,graphicx,graphics,amscd,amsfonts,url}

\begin{document}

\begin{flushright} 
March 2014 \\
OIQP-14-04
\end{flushright} 

\vspace{0.1cm}

\begin{Large}
\vspace{1cm}
\begin{center}
{\bf 
Multi-matrix models and Noncommutative Frobenius algebras obtained from symmetric groups 
and Brauer algebras \\
}
\end{center}
\end{Large}

\vspace{0.2cm}

\begin{center}
{\large Yusuke Kimura}

\quad 

Okayama Institute for Quantum Physics (OIQP), \\
Kyoyama 1-9-1, Kita-ku, Okayama, 700-0015, JAPAN \\
londonmileend \_at\_ gmail.com

\end{center}

\vspace{0.2cm}

\begin{abstract}

It has been understood that correlation functions of multi-trace operators 
in ${\cal N}=4$ SYM 
can be neatly computed 
using the group algebra of symmetric groups or walled Brauer algebras. 
On the other hand such algebras have been known to construct 2D 
topological field theories (TFTs). 
After reviewing the construction of 
2D TFTs based on symmetric groups,
we construct 2D TFTs
based on walled Brauer algebras. 
In the construction, the introduction of a dual basis manifests 
a similarity between the two theories.
We next construct 
a class of 2D field theories whose physical operators have the same symmetry 
as multi-trace operators constructed from some matrices. 
Such field theories correspond to
non-commutative Frobenius algebras. 
A matrix structure arises as a consequence of 
the noncommutativity. 
Correlation functions of the Gaussian complex multi-matrix models 
can be translated into 
correlation functions of the two-dimensional field theories.

\end{abstract}

\vspace{0.5cm}


\section{Introduction}

Motivated by the development of AdS/CFT correspondence \cite{9711200}, 
the study of ${\cal N}=4$ super Yang-Mills theory has attract lots of interests 
for these 15 years. 
The anomalous dimensions are of particular interest because they 
correspond to energies in the dual string theory side. 
If we restrict our attention to the planar limit, the problem to obtain 
the spectrum of anomalous dimensions   
is replaced with the conventional problem to 
diagonalise 
the Hamiltonian 
in an integrable system \cite{0212208,1012.3982}. 
Compared to the enormous development of the planar theory, 
we had only limited information on non-planar corrections. 
Recent development to study non-planar corrections, however, has brought us 
to obtain some concrete results of non-planar 
corrections \cite{0111222}-\cite{1301.1980}. 

Consider multi-trace gauge invariant operators 
constructed from a single complex matrix, relevant  
to the half-BPS sector
in ${\cal N}=4$ super Yang-Mills theory. 
They are classified by conjugacy classes of the symmetric group, 
and the two-point function  
can be expressed  
except the trivial space-time dependence as  
\begin{eqnarray}
\langle O_{[\tau]} O_{[\sigma]}\rangle 
&=&
\sum_{\rho \in S_n}N^n \delta_n(\Omega_n \tau \rho \sigma \rho^{-1}).
\label{2pt_intro}
\end{eqnarray}
The derivation will be given in section 2.  
It is interesting to find that the right-hand side is 
written completely in terms of symmetric group data, suggesting 
an effective role of the symmetric group in the problem. 
Behind the fact  
that the symmetric group shows up 
in the evaluation of the matrix integral, there is 
a mathematical structure that relates 
the general linear group and the symmetric group known as Schur-Weyl duality. 
It indeed plays a central role in 
our idea to employ
group representation theory 
in the recent development of non-planar physics. 
We will review it in section \ref{sec:N=4_review}, 
and some mathematical notions regarding Schur-Weyl duality 
are given in appendix \ref{SWdual_symmetric_Brauer}.    
By the way it has been known that symmetric groups give the description 
of coverings of two-dimensional Riemann surfaces, which 
played a central role in the 
string theoretic interpretation of  
the large $N$ expansion of two-dimensional Yang-Mills \cite{GrossTaylor,9402107,9411210}. 
This is an example that symmetric groups are used to describe 
two-dimensional field theories. 
In fact, 
if we ignore the $\Omega$ factor in the right-hand side of (\ref{2pt_intro}), 
the right-hand side is nothing but the two-point function of 
a two-dimensional topological field theory.  
In section 3 we give a review of the construction of 
two-dimensional topological field theories associated with symmetric groups. 
From these facts, it is expected to be fruitful to learn four-dimensional theories 
from two-dimensional theories by means of symmetric groups \cite{1301.1980}.  

Walled Brauer algebras \cite{Stembridge,Koike,Turaev,BCHLLS} 
are another convenient tool to organise 
multi-trace gauge invariant operators constructed from some matrices \cite{0709.2158,0807.3696}. 
In section \ref{topological_Brauer}, we will explicitly construct 
two-dimensional topological field theories based on walled Brauer algebras.  
The introduction of a dual basis manifests a similarity to the construction of 
topological field theories by symmetric groups. 
We will also show that correlation functions of the Brauer topological field theory 
can be decomposed into correlation functions of the topological field theories 
obtained from symmetric groups. 

The central part of this paper is, based on these connections, 
devoted to the study of 
two-dimensional field theories that are 
closely related to the description of multi-trace operators 
built from some scalar fields 
in ${\cal N}=4$ super Yang-Mills theory.
As will be reviewed in section \ref{sec:N=4_review}, when we organise gauge invariant operators 
that involve $p$ kinds of matrices in terms of symmetric groups or walled Brauer algebras,  
the conjugation under some permutation 
belonging to 
the symmetric group $S_{n_1}\times \cdots \times S_{n_p}$ 
is essentially 
important \cite{0709.2158,0711.0176,0801.2061,0805.3025,0807.3696,0910.2170,1206.4844}. 
In section \ref{sec:new_theory_subgroup} we will construct
two-dimensional field theories whose physical operators are characterised 
by the same conjugation.  
The idea of the construction is given in \cite{1301.1980}. 
It has been understood that two-dimensional topological quantum field theories 
are in  
one-to-one correspondence with commutative Frobenius algebras. 
What we describe in section \ref{sec:TFT_review} and \ref{topological_Brauer}
can be phrased mathematically that commutative Frobenius algebras are constructed from 
symmetric groups and walled Brauer algebras. 
By contrast, 
algebras playing a role in the theories we consider in section \ref{sec:new_theory_subgroup} 
are non-commutative Frobenius algebras. 
These theories have some properties owned by topological field theories 
such as that the two-point function is a projector. 
We will see that 
an effective matrix-like structure shows up as a consequence of the noncommutativity, 
where multiplicity 
indices on the restricted characters 
\cite{0701066,0709.2158,0801.2061,0807.3696} behave like matrix-indices.
We will also discuss the connection 
between correlation functions of the two-dimensional field theories and 
correlation functions of ${\cal N}=4$ super Yang-Mills theory 
in section \ref{N=4SYM_2pt_TFT}, hoping to illuminate a new geometric correspondence
between the two theories.  
In section \ref{ref:relations_counting}, 
we discuss the counting problem of multi-traces 
from the point of view of 
the two-dimensional field theories. 
Section \ref{discussions} is devoted to discussions. 
In some appendices we collect useful materials.


\section{Correlation functions in multi-matrix models}
\label{sec:N=4_review}
In this section, we will review the  
recent approach 
to compute exact finite $N$ 
correlation functions of multi-trace operators 
in ${\cal N}=4$ Super Yang-Mills theory.  
The mathematical background behind the approach is supplemented 
in appendix \ref{SWdual_symmetric_Brauer}. 

We consider the free field theory of scalar fields, and we use 
the complex notation to denote them by $X,Y,Z$ or $X_a$ ($a=1,2,3$). 
The two-point function is determined by conformal symmetry to 
take the following form
\begin{eqnarray}
\langle O_{i}(x)O_{j}(y)\rangle _{SYM}
=\frac{c_{ij}}{(x-y)^{n_i+n_j}}\delta_{n_i, n_j}, 
\end{eqnarray}
where $n_i$ is the number of fields involved in the operator $O_i$. 
General gauge invariant operators 
are given by a product of an arbitrary number of 
single trace operators built from 
the fundamental fields $X_a$ and $X_a^{\dagger}$. 
The non-trivial 
$N$-dependence of the correlator 
is encoded in 
$c_{ij}$, and it is obtained by solving 
the combinatorial problem of contractions of the free field 
captured by
the matrix integral with the Gaussian weight 
\cite{0205033}
\begin{eqnarray}
c_{ij}=
\langle O_{i}O_{j}\rangle 
:=
\int \prod_a [dX_a dX_a^{\dagger}] 
e^{-2 tr(X_aX_a^{\dagger})}
O_{i}O_{j},
\end{eqnarray}
where 
the measure is normalised to give 
$\langle  (X_a){}_{ij}(X_b^{\dagger}){}_{kl}
\rangle =
\delta_{il}\delta_{jk}\delta_{ab}$.

The evaluation of the matrix integral will be neatly performed 
if we introduce an appropriate algebra as a 
tool to organise the multi-trace structure. 
Let us first consider 
the half BPS chiral primary operators 
described by a single holomorphic matrix $X$ \cite{0111222,0205221}. 
Any multi-trace operators built from $n$ copies of $X$ are conveniently labelled by 
an element of the symmetric group $S_n$ as
\begin{eqnarray}
tr_{n}(\sigma X^{\otimes n})=
X_{i_1}^{i_{\sigma(1)}}X_{i_2}^{i_{\sigma(2)}}\cdots
X_{i_n}^{i_{\sigma(n)}} \quad (\sigma\in S_n).
\label{one-matrix_symmetric}
\end{eqnarray}
Here $tr_n$ is a trace over the tensor space $V^{\otimes n}$, 
where $X$ is regarded as a linear map on $V$. 
Because 
two elements that are conjugate each other give the same multi-trace
\begin{eqnarray}
tr_{n}(\sigma X^{\otimes n})=
tr_{n}(\rho\sigma \rho^{-1} X^{\otimes n})
\quad (\rho\in S_n), 
\label{equivalence_class_one_matrix}
\end{eqnarray}
the number of independent multi-traces is not equal to the 
number of elements in the symmetric group $S_n$.
Taking the equivalence relation into account, 
the multi-trace operators are correctly classified by conjugacy classes. 
The matrix integral of 
the two-point function can be evaluated as
\begin{eqnarray}
\langle tr_{n}(\tau X^{\dagger \otimes n})tr_{n}(\sigma X^{\otimes n})\rangle 
&=&\sum_{\rho\in S_n}tr_n(\tau \rho \sigma \rho^{-1})
\nonumber \\
&=&\sum_{\rho\in S_n}\sum_{R\vdash n}t_R\chi_R(\tau \rho \sigma \rho^{-1})
\nonumber \\
&=&
\sum_{h\in S_n}N^n \delta_n(\Omega_n \tau \rho \sigma \rho^{-1}), 
\label{derive_twopt_single}
\end{eqnarray}
At the first equality the Wick-contractions are expressed by 
elements in $S_n$ \cite{0111222}. 
The second and third step result from the Schur-Weyl duality, as is explained in 
(\ref{swdual_trace}) - (\ref{tr_n_tau_evaluate}). 
The dimension of 
an irreducible representation $R$ of the symmetric group 
is denoted by $d_R$, while 
the dimension of 
an irreducible representation $R$ of the $SU(N)$ group
is denoted by $t_R$. 
The $\Omega_n$ is a specific central element in the group algebra of 
the symmetric group,
\begin{eqnarray}
\Omega_n=\sum_{\sigma\in S_n}\sigma N^{C_{\sigma}-n},
\end{eqnarray}
where 
$C_{\sigma}$ is the number of cycles in the permutation $\sigma$. 
The last equality of (\ref{derive_twopt_single}) 
is valid for $N$ that is larger than $n$. 
The two-point functions can be diagonalised by the basis change 
of (\ref{projector_S_n_single}) as \cite{0111222} 
\begin{eqnarray}
\langle tr_{n}(p_R X^{\dagger \otimes n})tr_{n}(p_S X^{\otimes n})\rangle 
&=& n!d_R t_R\delta_{RS}. 
\end{eqnarray}
The diagonal basis is labelled by 
a Young diagram with $n$ boxes.

The idea to label multi-trace operators in terms of 
an element in the group algebra of the 
symmetric group 
can be applied to 
operators described by some
kinds of matrices \cite{0411205,0711.0176,0801.2061}. 
For a multi-trace 
constructed from  
$m$ copies of $X$ and $n$ copies of $Y$, using $\sigma\in S_{m+n}$ we have 
\begin{eqnarray}
tr_{m+n}(\sigma X^{\otimes m}\otimes Y^{\otimes n})=
X_{i_1}^{i_{\sigma(1)}}\cdots
X_{i_m}^{i_{\sigma(m)}}
Y_{i_{m+1}}^{i_{\sigma(m+1)}}\cdots
Y_{i_{m+n}}^{i_{\sigma(m+n)}}. 
\label{two-matrix_symmetric}
\end{eqnarray}
The difference from (\ref{one-matrix_symmetric}) 
is that    
equivalence classes for 
(\ref{two-matrix_symmetric}) are 
characterised by the 
equivalence relation determined by the subgroup
\begin{eqnarray}
tr_{m+n}(\sigma X^{\otimes m}\otimes Y^{\otimes n})=
tr_{m+n}(h\sigma h^{-1} X^{\otimes m}\otimes Y^{\otimes n})
\quad (h\in S_m\times S_n).
\label{equivalence_class_two_matrix_symmetric}
\end{eqnarray}
In this description the subgroup $H=S_m\times S_n$ plays a role. 
Two-point function can be evaluated 
as 
\begin{eqnarray}
\langle tr_{m+n}(\tau X^{\dagger \otimes m}\otimes Y^{\dagger \otimes n})
tr_{m+n}(\sigma X^{\otimes m}\otimes Y^{\otimes n})\rangle 
&=&\sum_{h\in S_m \times S_n}tr_{m+n}(\tau h \sigma h^{-1})
\nonumber \\
&=&
\sum_{h\in S_m \times S_n}N^{m+n} \delta_{m+n}(\Omega_{m+n} \tau h \sigma h^{-1}). 
\label{two-point_two-matrix_symmetric}
\end{eqnarray}
We emphasise that the free field Wick-contractions are expressed by 
elements of 
the subgroup $S_m\times S_n$. 
A diagonal two-point function can be obtained 
by a change of basis in (\ref{intertwiner_symmetric_group}) \cite{0801.2061},
\begin{eqnarray}
\langle tr_{m+n}(P^R_{A,\mu\nu} X^{\dagger \otimes m}\otimes Y^{\dagger \otimes n})
tr_{m+n}(P^S_{A^{\prime},\mu^{\prime}\nu^{\prime}} X^{\otimes m}
\otimes Y^{\otimes n})\rangle 
&=& m!n!d_A t_R\delta_{RS}\delta_{AA^{\prime}}
\delta_{\mu\nu^{\prime}}
\delta_{\nu\mu^{\prime}}.
\label{restricted_schur_two-point}
\end{eqnarray}
The diagonal operators are labelled by a set of 
three Young diagrams and two multiplicity labels. 
Another diagonal basis is described in \cite{0711.0176}.

We have another way to label multi-traces 
constructed from some fields 
using walled Brauer algebras \cite{0709.2158,0807.3696}. 
Walled Brauer algebras can be introduced as a Schur-Weyl dual to 
the $GL(N)$ groups (see (\ref{SWdual_Brauer})). 
Mainly consider multi-trace operators constructed from 
$m$ copies of $X$ and $n$ copies of $Y$, and let 
$B_N(m,n)$ be the walled Brauer algebra 
relevant for the description of such operators. 
The Brauer algebra contains the group algebra of $S_m\times S_n$ as a subalgebra. 
Gauge invariant operators are constructed 
by regarding $X^{\otimes m}\otimes Y^{T\otimes n}$ as operators 
acting on the space 
$V^{\otimes m}\otimes \bar{V}^{\otimes n}$, followed by 
the action of an element of the Brauer algebra and taking a trace:
\begin{eqnarray}
tr_{m,n}(b X^{\otimes m}\otimes Y^{T\otimes n}) 
\quad (b\in B_N(m,n)).  
\label{two-matrix_brauer}
\end{eqnarray}
The equivalence relation is very similar to 
(\ref{equivalence_class_two_matrix_symmetric}),
\begin{eqnarray}
tr_{m,n}(b X^{\otimes m}\otimes Y^{T\otimes n})=
tr_{m,n}(hb h^{-1} X^{\otimes m}\otimes Y^{T\otimes n})
\quad (h\in S_m\times S_n).
\label{equivalence_class_two_matrix_Brauer}
\end{eqnarray}
By expressing free field Wick-contractions in terms of 
elements in $H$, 
the two-point function is computed to give 
\begin{eqnarray}
\langle tr_{m,n}(b X^{\dagger \otimes m}\otimes Y^{T\dagger \otimes n})
tr_{m,n}(c X^{\otimes m}\otimes Y^{T\otimes n})\rangle 
&=&\sum_{h\in S_m \times S_n}tr_{m,n}(b h c h^{-1})
\nonumber \\
&=&\sum_{h\in S_m \times S_n}
\sum_{\gamma}t_\gamma \chi^{\gamma}(b h c h^{-1}),
\label{two-point_Brauer}
\end{eqnarray}
where $b,c$ are elements of the walled Brauer algebra.
The dimension of 
an irreducible representation $\gamma$ of the walled Brauer algebra 
is denoted by $d_{\gamma}$, while 
the dimension of 
an irreducible representation $\gamma$ of the $GL(N)$ group
is denoted by $t_{\gamma}$.
The last equality in 
(\ref{two-point_Brauer}) is a consequence of the Schur-Weyl duality 
(\ref{SWdual_Brauer}). 
By taking the linear combination in  
(\ref{explicit_form_Q-operator}), we will obtain 
diagonal two-point functions \cite{0709.2158},
\begin{eqnarray}
\langle tr_{m,n}(Q^{\gamma}_{A,\mu\nu} X^{\dagger \otimes m}\otimes Y^{T\dagger \otimes n})
tr_{m,n}(Q^{\gamma^{\prime}}_{A^{\prime},\mu^{\prime}\nu^{\prime}} 
X^{\otimes m}\otimes Y^{T \otimes n})\rangle 
= m!n!d_A t_{\gamma}\delta_{\gamma\gamma^{\prime}}\delta_{AA^{\prime}}
\delta_{\mu\nu^{\prime}}\delta_{\nu\mu^{\prime}}.
\end{eqnarray}
Walled Brauer algebras can also be used to describe 
multi-trace operators constructed from 
more than two kinds of matrices \cite{0910.2170,1206.4844}. 

\quad 

Before closing this section, we will introduce some symbols 
to denote the equivalence classes characterising the multi-matrix structures.
For a fixed element $\sigma \in S_n$ 
the following sum 
over a complete basis of $S_n$ gives a central element 
\begin{eqnarray}
[\sigma]=\frac{1}{n!}\sum_{\rho \in S_n}\rho\sigma \rho^{-1} \quad 
(\sigma \in S_n). 
\label{conjugacy_class_symbol}
\end{eqnarray}
In symmetric group $S_{m+n}$ the sum over a 
complete basis of $S_{m}\times S_n$ gives an 
element which commutes with any elements in $S_{m}\times S_n$, 
\begin{eqnarray}
[\sigma]_H=\frac{1}{m!n!}\sum_{h \in H}h\sigma h^{-1} \quad 
(\sigma \in S_{m+n}),
\label{conjugacy_class_subgroup_symbol}
\end{eqnarray}
where $H=S_m\times S_n$.
Similarly an element in the Brauer algebra $B_N(m,n)$ 
which commutes with any elements in $S_{m}\times S_n$ 
can be constructed by 
\begin{eqnarray}
[b]_H=\frac{1}{m!n!}\sum_{h \in H}hb h^{-1} \quad 
(b \in B_N(m,n)). 
\label{conjugacy_class_subgroup_Brauer_symbol}
\end{eqnarray}
The number of the equivalence classes 
coincides with the number of independent multi-matrix operators at large $N$ 
(see also (\ref{counting_new_2D_symmetric}), (\ref{counting_new_2D_Brauer}) and 
(\ref{matching_counting})).
If $N$ is small compared to the number of fields involved in the multi-traces, 
multi-traces are in general linearly dependent. 
Linearly independent multi-traces are given by 
the Young diagram basis \cite{0111222,0709.2158,0711.0176,0801.2061,0107119}.

With these new notations, the two-point functions are rewritten as 
\begin{eqnarray}
&&
\langle tr_{n}([\tau] X^{\dagger \otimes n})tr_{n}([\sigma] X^{\otimes n})\rangle 
=
n!N^n \delta_n(\Omega_n [\tau ][\sigma])
\nonumber \\
&&
\langle tr_{m+n}([\tau]_H X^{\dagger \otimes m}\otimes Y^{\dagger \otimes n})
tr_{m+n}([\sigma]_H X^{\otimes m}\otimes Y^{\otimes n})\rangle 
=m!n!
N^{m+n} \delta_{m+n}(\Omega_{m+n} [\tau ]_H[\sigma]_H)
\nonumber \\
&&
\langle tr_{m,n}([b]_H X^{\dagger \otimes m}\otimes Y^{T\dagger \otimes n})
tr_{m,n}([c]_H X^{\otimes m}\otimes Y^{T\otimes n})\rangle 
=m!n!
\sum_{\gamma}t_\gamma \chi^{\gamma}([b]_H[c]_H). 
\label{2pt_functions_new_notation}
\end{eqnarray}
We will reconsider the meaning of these equations in section \ref{discussions}.


\section{2D topological field theories and semisimple algebras}
\label{sec:TFT_review}
In the previous section we have reviewed that employing 
the group algebra of the symmetric group and the walled Brauer algebra 
is very effective for labelling multi-trace gauge invariant operators 
in the multi-matrix models. 
On the other hand 
it has been known that 
two-dimensional 
topological quantum
field theories can be defined from
such algebras. 
This fact would bring up a new aspect on the role of the algebras, suggesting 
a new connection with
two-dimensional field theories. 
In this section we will review 
the construction of two-dimensional topological quantum field theories.  
For our purpose it is convenient to define them 
in terms of lattice \cite{FHK,9205031,9305080}, where it was shown that 
semisimple algebras have a one-to-one correspondence with two-dimensional 
topological quantum field theories.

Consider the triangulation of 
a two-dimensional compact orientable surface.  
The structure of the surface is encoded in the way of gluing the triangles. 
To each edge we assign a colour index $i$, 
and  
to each triangle with edges labelled by $i,j,k$, 
we assign a complex number $C_{ijk}$. 
We assume that $C_{ijk}$ is invariant under cyclic permutations of the colour indices
\begin{eqnarray}
C_{ijk}=C_{jki}=C_{kij}, 
\end{eqnarray}
but no relation is imposed between $C_{ijk}$ and the orientation-reversed 
object $C_{ikj}$. 
Two adjacent edges are identified by introducing a gluing operator $g^{ij}$, 
which is assumed to be symmetric $g^{ij}=g^{ji}$ and to have the inverse $g_{ij}$. 
In the dual diagrams $g^{ij}$ and $C_{ikj}$ correspond to 
the propagator and the three-point vertex respectively.  
The partition function of a triangulated surface
is given by the product of complex numbers $C_{ijk}$ 
assigned to each triangle 
with each edge glued by the gluing operator $g^{ij}$ and the summation over all 
possible triangulations.  

\begin{figure}[t]
\begin{center}
 \includegraphics[scale=0.6]{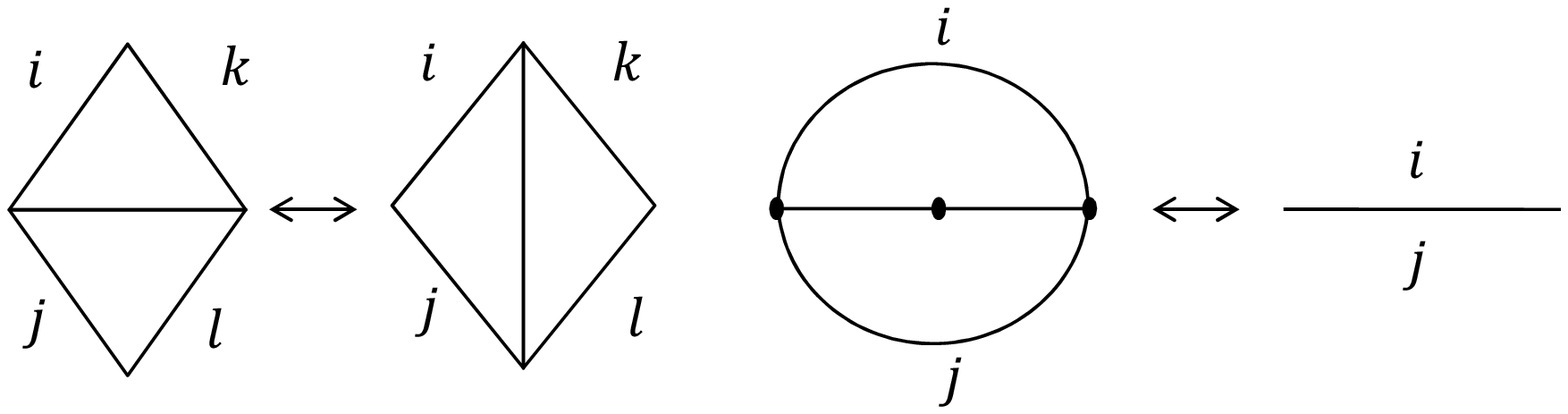}
\caption{
Left - the 2-2 move, Right - the bubble move
} 
\label{fig:topological_moves}
\end{center}
\end{figure}

Topological quantum field theories are characterised by 
the invariance under local deformations of the background. 
In the lattice construction, topological models are constructed by imposing the invariance 
of partition functions under any local change of the triangulations. 
It is known that two basic moves are sufficient to 
generate all topologically equivalent 
triangulations. 
We will use so called bubble move and 2-2 move (see figure \ref{fig:topological_moves}).
From 
the invariance of 
partition functions under 
the bubble moves,  
we obtain 
\begin{eqnarray}
C_{ik}{}^lC_{jl}{}^k=g_{ij},
\label{CC=g}
\end{eqnarray}
where 
\begin{eqnarray}
C_{ij}{}^k=C_{ijl}g^{lk}.
\label{C=Cg}
\end{eqnarray}
We use the usual summation convention that repeated indices are summed. 
From 
the invariance of 
partition functions under the 
2-2 moves,  
we have 
\begin{eqnarray}
C_{ij}{}^p C_{pk}{}^l=C_{ip}{}^lC_{jk}{}^p . 
\label{CC=CC}
\end{eqnarray}
These two conditions are solved by introducing 
a semisimple associative algebra \cite{FHK,9205031}. 
Let a basis and the structure constant of the algebra be 
$\phi_{i}$ and $C_{ij}{}^k$, that is, $\phi_{i}\phi_{j}=C_{ij}{}^k\phi_{k}$.  
It is easy to find that the condition 
(\ref{CC=CC}) is derived from the associativity of the algebra; 
$(\phi_{i}\phi_{j})\phi_{k}=\phi_{i}(\phi_{j}\phi_{k})$. 
The condition (\ref{CC=g}) indicates that 
we can define 
a nondegenerate metric $g_{ij}$ given in the equation. 
With these conditions, partition functions 
only depend on the genus of the manifold.

In order to make the construction more concrete, 
we will 
consider the group algebra of symmetric group $S_n$ as an example of 
semisimple algebras \cite{FHK}. 
Introducing 
the regular representation, the structure constant is given by
\begin{eqnarray}
C_{ij}{}^{k}
=\frac{1}{n!}tr^{(r)}(\sigma_k^{-1} \sigma_i\sigma_j),
\end{eqnarray}
where $tr^{(r)}$ is the trace of the regular representation. 
Defining the delta function 
over the group algebra of $S_n$
by $\delta_n(\sigma)=1$ if $\sigma=1$ and 
$0$ otherwise, the trace of the regular representation can be expressed by 
\begin{eqnarray}
tr^{(r)}(\sigma)
=n!\delta_n(\sigma). 
\end{eqnarray}
It is also convenient to consider the expansion 
in terms of 
the character of the symmetric group as  
 \begin{eqnarray}
tr^{(r)}(\sigma)
=
\sum_{R\vdash n}d_{R}\chi_{R}(\sigma), 
\end{eqnarray}
where $R\vdash n$ means that $R$ is a Young diagram with $n$ boxes. 
From this formula 
we can verify $tr^{(r)}(1)=\sum_{R\vdash n}(d_R)^2$, where 
$d_R$ is the dimension of 
an irreducible representation $R$ of $S_n$.

From (\ref{CC=g}) and (\ref{C=Cg}) we determine 
\begin{eqnarray}
&&
g_{ij}=tr^{(r)}(\sigma_i\sigma_j )=n!\delta_n(\sigma_i\sigma_j ) 
\nonumber 
\\
&&
C_{ijk}
=tr^{(r)}(\sigma_i\sigma_j \sigma_k)=
n!\delta_n(\sigma_i\sigma_j \sigma_k), 
\label{metric_C_symmetric_def}
\end{eqnarray}
and 
\begin{eqnarray}
g^{ij}=\frac{1}{(n!)^2}
tr^{(r)}(\sigma_i^{-1}\sigma_j ^{-1})
=
\frac{1}{n!}\delta(\sigma_i^{-1}\sigma_j^{-1}). 
\label{inverse_g^ij}
\end{eqnarray}
Let us now introduce a dual basis defined by 
\begin{eqnarray}
\sigma^i:=g^{ij}\sigma_j=\frac{1}{n!}\sigma_i^{-1}, 
\label{Brauer_dual_basis}
\end{eqnarray}
which satisfies 
\begin{eqnarray}
\sum_i\sigma^i\sigma_i=1. 
\end{eqnarray}
If we use the dual basis, we do not have an extra factor arising 
from raising or lowering indices
\begin{eqnarray}
&&g^{ij}=tr^{(r)}(\sigma^i\sigma^j)=n!\delta(\sigma^i\sigma^j),
\nonumber \\
&&C_{ij}{}^{k}
=tr^{(r)}(\sigma_i\sigma_j\sigma^k)=
n!\delta( \sigma_i \sigma_j\sigma^k). 
\end{eqnarray}

\begin{figure}[t]
\begin{center}
 \includegraphics[scale=0.6]{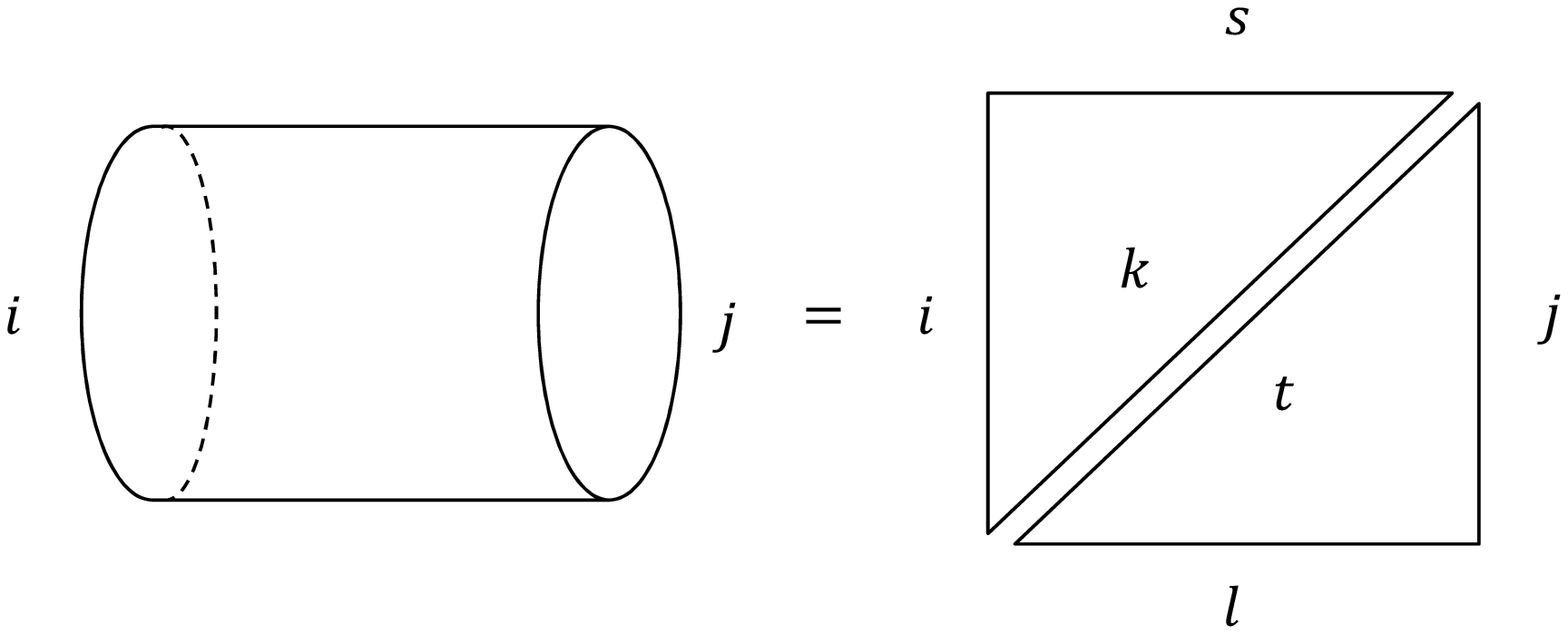}
\caption{
a triangulation of the cylinder with two boundaries $i,j$ : 
the edges $k$ and $t$ and the edges $s$ and $l$ 
being glued by the operator $g^{ij}$ as in (\ref{prescription_eta_symmetric})
} 
\label{fig:cylinder}
\end{center}
\end{figure}

\begin{figure}[t]
\begin{center}
 \includegraphics[scale=0.6]{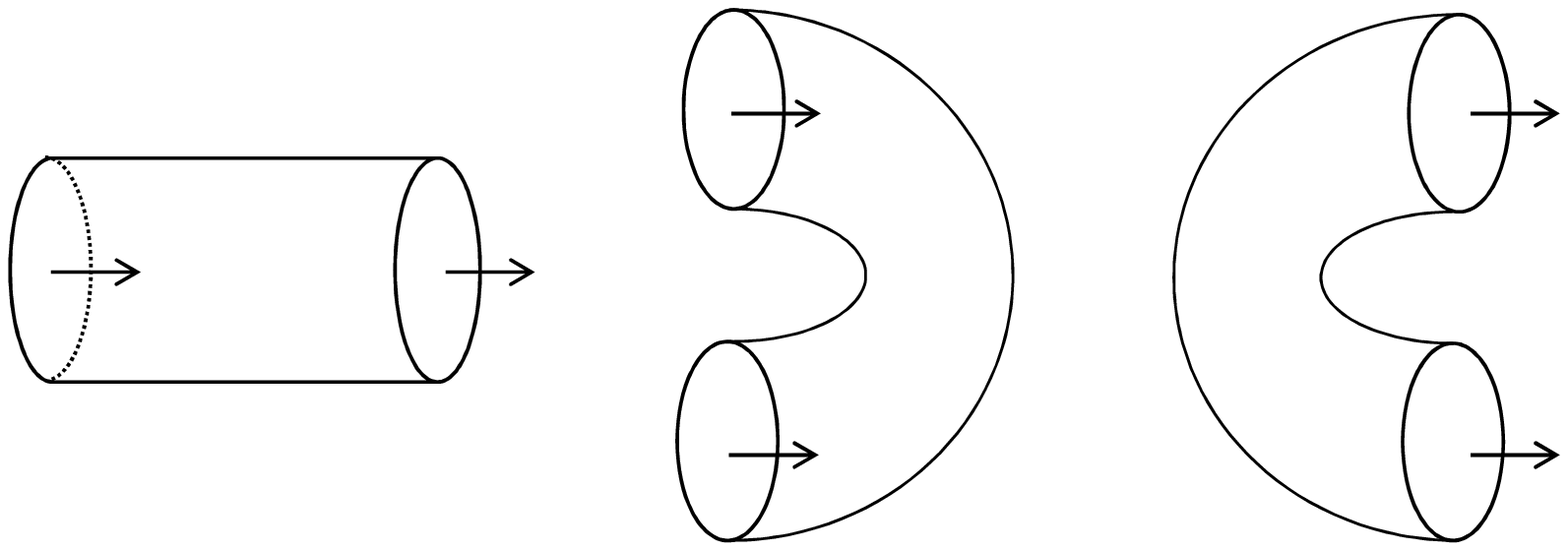}
\caption{
three oriented cylinders, being drawn so that all arrows are turning to the right
} 
\label{fig:oriented_cylinder}
\end{center}
\end{figure}

We next construct 
the two-point function on a sphere.
A simple triangulation shown in figure \ref{fig:cylinder}
leads to 
\begin{eqnarray}
\eta_{ij}=C_{iks}C_{ljt}g^{sl}g^{kt}
=C_{ik}{}^{l}C_{lj}{}^{k}. 
\label{prescription_eta_symmetric}
\end{eqnarray}
We can show that 
\begin{eqnarray}
&&
\eta_{ij}
=\sum_{R\vdash n}\chi^{R}(\sigma_i)\chi^{R}(\sigma_j)
\nonumber \\
&&
\eta^{ ij}
=\sum_{R\vdash n}\chi^{R}(\sigma^i)\chi^{R}(\sigma^j)
\nonumber \\
&&
\eta_{i}{}^{j}
=\sum_{R\vdash n}\chi^{R}(\sigma_i)\chi^{R}(\sigma^j), 
\label{projector-symmetric}
\end{eqnarray}
or we have 
\begin{eqnarray}
&&\eta_{ij}=\sum_{k}tr^{(r)}(\sigma_i\sigma_k \sigma_j \sigma^k)
=tr^{(r)}([\sigma_i][\sigma_j])
\nonumber \\
&&\eta^{ij}=\sum_{k}tr^{(r)}(\sigma^i\sigma_k \sigma^j \sigma^k)
=tr^{(r)}([\sigma^i][\sigma^j])
\nonumber \\
&&\eta_{i}{}^{j}=\sum_{k}tr^{(r)}(\sigma_i\sigma_k \sigma^j \sigma^k)
=tr^{(r)}([\sigma_i][\sigma^j]). 
\end{eqnarray}
One important property is that this is a projection operator,
$\eta^2=\eta$. More explicitly, we have
\begin{eqnarray}
\eta_{ij}\eta^{jk}=\eta_i{}^k, \quad 
\eta_{i}{}^{j}\eta_{j}{}^{k}=\eta_i{}^k, \quad 
\eta_{ij}\eta^{j}{}_{k}=\eta_{ik},
\label{explicit_projector_relation}
\end{eqnarray}
and so on. 
The fact that a cylinder is a projection operator is a general property of 
topological field theories \cite{Atiyah}. 
Lower indices and upper indices correspond to different orientations. 
The difference comes from the difference between 
the basis and the dual basis.  
In order to distinguish them it is convenient to to call the two kinds of boundaries 
{\it in-boundaries} and {\it out-boundaries}.  
We can draw three kinds of cylinders as in 
figure \ref{fig:oriented_cylinder}, corresponding to 
$\eta_i{}^j$, $\eta_{ij}$ and $\eta^{ij}$.  
The readers can find the diagrammatic meaning of the relation 
(\ref{explicit_projector_relation}) using the oriented cylinders
in figure \ref{fig:oriented_cylinder}. 
In this theory, 
the difference is just the factor of $1/n!$, 
\begin{eqnarray}
\chi^R(\sigma^i)=\frac{1}{n!}\chi^R(\sigma_i^{-1})=\frac{1}{n!}\chi^R(\sigma_i),
\end{eqnarray}
because $\sigma^{-1}$ is conjugate to $\sigma$. 
The difference 
between different orientations 
will be more important in theories we will consider in what follows.

The projection operator determines the space of physical operators \cite{FHK}. 
It can be shown that 
the $\eta$ is the projection operator from the algebra onto 
the centre of the algebra,
\begin{eqnarray}
\eta_{i}{}^{j}\sigma_j
=
\sum_{R \vdash n}\frac{1}{d_{R}}
\chi^{R}(\sigma_i)
p_{R}=[\sigma_i], 
\label{projection_physical_eta}
\end{eqnarray}
where $p_R$ is a central element given in (\ref{projector_S_n_single}), 
and 
$[\sigma]$ is defined in (\ref{conjugacy_class_symbol}).
In other words,
physical states are invariant under time translation,
\begin{eqnarray}
\eta_{i}{}^{j}[\sigma_j]=[\sigma_i].
\end{eqnarray}
We thus have a one-to-one correspondence between 
central elements and 
physical states. 
The number of the physical states is the number of 
conjugacy classes.

Because
the cylinder is 
a projection operator onto the physical Hilbert space, the torus, 
which is obtained by gluing two boundaries of the cylinder, gives
the dimension of the vector space \cite{Atiyah,Dijkgraaf:1989pz},
\begin{eqnarray}
Z_{G=1}
=\eta_i{}^i=\sum_{R\vdash n}1.
\end{eqnarray}
This counts the number of Young diagrams built from $n$ boxes. 
It is equivalent to the number of multi-matrices built 
from $n$ copies of a matrix at large $N$.

The three-point function $N_{ijk}$ is obtained 
from 
the structure constant of the operator product of physical states,
\begin{eqnarray}
[\sigma_i][\sigma_j]
=N_{ij}{}^k
[\sigma_k].
\end{eqnarray}
From this we find 
\begin{eqnarray}
N_{ijk}=
tr^{(r)}([\sigma_i][\sigma_j][\sigma_k])
=
\sum_{R\vdash n}
\frac{1}{d_{R}}
\chi^{R}(\sigma_i)\chi^{R}(\sigma_j)\chi^{R}(\sigma_k)
\label{three_point_etaTFT}
\end{eqnarray}
and 
\begin{eqnarray}
N_{ij}{}^k
=tr^{(r)}([\sigma_i][\sigma_j][\sigma^k])
=
\sum_{R\vdash n}
\frac{1}{d_{R}}
\chi^{R}(\sigma_i)\chi^{R}(\sigma_j)\chi^{R}(\sigma^k),
\end{eqnarray}
where $N_{ijk}=N_{ik}{}^{l}\eta_{lk}$. 
Because this operator algebra is commutative
\begin{eqnarray}
[\sigma_i][\sigma_j]=[\sigma_j][\sigma_i], 
\end{eqnarray}
we have 
$N_{ij}{}^k=N_{ji}{}^k$.
The associativity 
$([\sigma_i][\sigma_j])[\sigma_k]=[\sigma_i]([\sigma_j][\sigma_k])$ 
gives 
\begin{eqnarray}
N_{ij}{}^k N_{kl}{}^n=N_{ik}{}^n N_{jl}{}^k. 
\end{eqnarray}
From $[\sigma_i]=\eta_i{}^p\sigma_p$, 
we find that 
the cylinder and the three-holed sphere 
are also obtained by acting with the projection operator 
on $g_{ij}$ and $C_{ijk}$ 
as
\begin{eqnarray}
&&
\eta_{k}{}^{i}\eta_{l}{}^{j}g_{ij}=\eta_{kl}
\nonumber \\
&&
\eta_{l}{}^{i}\eta_{m}{}^{j}\eta_{n}{}^{k}C_{ijk}=N_{lmn}.
\label{physical_two-point_three-point}
\end{eqnarray}
These are also obtained by a simple triangulation \cite{FHK}. 

\begin{figure}[t]
\begin{center}
 \includegraphics[scale=0.6]{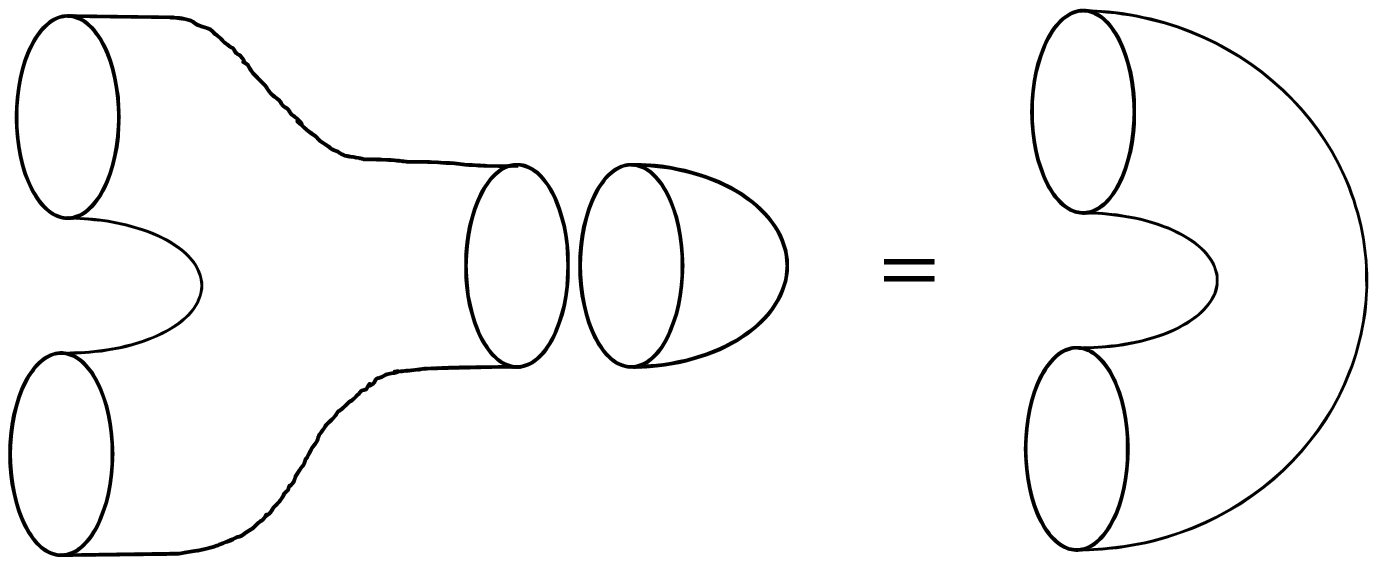}
\caption{
drawing of $N_{ij}{}^k\eta_{k0}=\eta_{ij}$
} 
\label{fig:Frobenius_Form}
\end{center}
\end{figure}

An associative algebra with a non-degenerate inner product is called 
Frobenius algebra.\footnote{
Important remarks are found in p.98 of \cite{Kock}. 
A Frobenius algebra is defined by 
providing with its Frobenius structure 
(Frobenius pairing or Frobenius form). 
If we choose two different Frobenius structures in an algebra, 
we will obtain two different Frobenius algebras. Being a Frobenius algebra is 
not a property of an algebra, but it is a structure of the algebra.
}
It has been understood that 
Frobenius algebras play a role in the algebraic and axiomatic formulation of 
topological quantum field theories, 
and 
commutative Frobenius algebras are in one-to-one correspondence 
with two-dimensional 
topological quantum 
field theories \cite{Abrams,Kock}.
In the example, the algebra of $[\sigma_i]$ with the metric $\eta_{ij}$ 
is a commutative Frobenius algebra. 
The $\eta_{ij}$ is called Frobenius pairing.  
The pairing also determines a linear map from the vector space 
to the ground field,
\begin{eqnarray}
\eta_{0i}=
\sum_{R\vdash n} d_R  \chi_R (\sigma_i)=n!\delta(\sigma_i) ,
\end{eqnarray}
where $\sigma_0=1$. 
This is called Frobenius form. Formally it is the sphere with a hole. 
The three objects, $\eta_{ij}$, $N_{ijk}$ and $\eta_{i0}$ can be 
basic building blocks to build up all surfaces. 
There is a relation among them, 
$N_{ij}{}^k\eta_{k0}=\eta_{ij}$ (see figure \ref{fig:Frobenius_Form}). 

All partition functions are built up from 
the building blocks 
$\eta_{ij}$ and $N_{ijk}$.
The partition function of a Riemann surface of genus $G$ is 
\begin{eqnarray}
Z_{G}
=\sum_{R \vdash n}\frac{1}{(d_{R}){}^{2G-2}}. 
\label{partition_function_symmetric_G}
\end{eqnarray}
The correlation function on 
the manifold of 
genus $G$ with $B_1$ in-boundaries and $B_2$ out-boundaries is given by  
\begin{eqnarray}
Z_{G,B_1,B_2}
=\sum_{R \vdash n}\frac{1}{(d_{R}){}^{2G+B_1+B_2-2}}
\chi^{R}(\sigma_{i_1})\cdots
\chi^{R}(\sigma_{i_{B_1}})\chi^{R}(\sigma^{j_1}) \cdots \chi^{R}(\sigma^{j_{B_2}}). 
\end{eqnarray}
The repeated use of (\ref{character_sum_sigma_more}) and 
(\ref{character_combine_useful}) 
helps us to find an equivalent form 
\begin{eqnarray}
Z_{G,B_1,B_2}
=
\sum_{k,l}
tr^{(r)}\left(
(\rho_{k_1}\tau_{l_1}\rho^{k_1}\tau^{l_1})
\cdots 
(\rho_{k_G}\tau_{l_G}\rho^{k_G}\tau^{l_G})
[\sigma_{i_1}] 
\cdots 
[\sigma_{i_{B_1}} ][\sigma^{j_1}] 
\cdots 
[\sigma^{j_{B_2}} ]
\right).
\end{eqnarray} 
We note that $\sum_{k,l}\rho_{k}\tau_{l}\rho^{k}\tau^{l}$ and 
$[\sigma]$ are central elements 
of the group algebra of the symmetric group, so their positions inside the trace 
are irrelevant. 
The form shows that the correlation functions are also obtained 
by representing the surface as a polygone with edges properly identified \cite{Witten:1991we}.

It is possible to choose a diagonal basis 
so that the operator product is given by 
$O_{\alpha}O_{\beta}=\delta_{\alpha\beta}O_{\alpha}$. 
This is realised by 
\begin{eqnarray}
p_{R}=\frac{d_{R}}{n!}\sum_{\sigma\in S_{n}}\chi_{R}(\sigma^{-1})[\sigma]
=d_{R}\sum_{i}\chi_{R}(\sigma^i)[\sigma_i]. 
\end{eqnarray}
The role of $p_R$ is explained around (\ref{projector_S_n_single}).
In this basis, $\eta_R{}^S=\delta_R{}^S$.


\section{2D topological field theories obtained from walled Brauer algebras}
\label{topological_Brauer}

\subsection{Construction of Brauer topological field theory}
In this section we present 
the topological lattice field theory obtained 
from the walled Brauer algebra $B_N(m,n)$, following the construction 
in the previous section. 
Walled Brauer algebras have a parameter $N$, which is identified with 
the matrix size $N$ when we use them to organise the multi-trace structure of 
$N\times N$ matrices. 
We will assume that $N$ is large enough 
so that $m+n<N$ is satisfied. 
This large $N$ condition secures the semisimplicity of the algebra,
which we need to 
construct a non-degenerate metric. 
In this construction 
we will use the idea presented in \cite{Ram_thesis} that 
general semisimple algebras can be analogously 
introduced to finite groups.  

Let a basis of the algebra be $b_i$. 
We first define a dual basis $b_i^{\ast}$ 
with respect to the bilinear form given by the trace $tr_{m,n}$ 
over the mixed tensor space $V^{\otimes m}\otimes \bar{V}^{\otimes n}$ 
in (\ref{SWdual_Brauer}), 
\begin{eqnarray}
tr_{m,n}(b_ib_j^{\ast})=\delta_{i}{}^{j}. 
\label{def_b_ast}
\end{eqnarray}
This dual basis was exploited in \cite{0709.2158}. 

The trace of 
the regular representation is introduced by 
\begin{eqnarray}
tr^{(r)}(b)=\sum_{i}tr_{m,n}(b_i^{\ast}bb_i)=
\sum_{\gamma}d^{\gamma}\chi^{\gamma}(b), 
\end{eqnarray}
where (\ref{tracemn_Brauer}) and 
(\ref{orthogonality_rep_brauer}) are
helpful to confirm the second equality, and 
$\gamma$ are irreducible representations of the walled Brauer algebra. 
An irreducible representation $\gamma$ is labelled by a bi-partition 
(see below equation (\ref{SWdual_Brauer}).)
Here $d^{\gamma}$ is the dimension of an irreducible representation $\gamma$
of the walled Brauer algebra, while $t^{\gamma}$ will be that of the $GL(N)$ group. 
The appearance of 
the $GL(N)$ group is a consequence of the Schur-Weyl duality.
The multiplication of the algebra, $b_ib_j=C_{ij}{}^{k}b_k$, is determined by
\begin{eqnarray}
C_{ij}{}^{k}=
tr_{m,n}(b_k^{\ast}b_ib_j). 
\end{eqnarray}
Orthogonality relations in appendix 
\ref{sec:formula_orthogonality} will give 
\begin{eqnarray}
&&
g_{ij}=tr^{(r)}(b_ib_j ), 
\\
&&
C_{ijk}=tr^{(r)}(b_ib_j b_k), 
\end{eqnarray}
and 
\begin{eqnarray}
g^{ij}=
\sum_{\gamma}
\frac{t^{\gamma}{}^2}{d^{\gamma}}
\chi^{\gamma}(b_i^{\ast}b_j^{\ast})
\label{g^ij}.
\end{eqnarray}

Let us now 
introduce another dual basis by 
\begin{eqnarray}
b^i=g^{ij}b_j, 
\label{def_of_b^i}
\end{eqnarray}
and we can confirm the following equation 
\begin{eqnarray}
\sum_i b^ib_i=1.
\end{eqnarray}
The following equation is helpful to relate the first dual basis 
$b_i^{\ast}$ and the second dual basis $b^i$,
\begin{eqnarray}
D^{\gamma}(b^i)=
g^{ij}D^{\gamma}(b_j)=\frac{t^{\gamma}}{d^{\gamma}}D^{\gamma}(b_i^{\ast}) ,
\end{eqnarray}
where $D^{\gamma}(b)$ is the representation 
matrix of $b$ in $\gamma$. 
From this, 
we find that $b^i$ is a dual basis with respect to the bilinear form determined by 
the trace of the 
regular representation, 
\begin{eqnarray}
tr^{(r)}(b_ib^j)=\delta_i{}^j.
\label{bast_b^i}
\end{eqnarray}
Defining a dual basis allows us to obtain the representation theory of general semisimple 
algebras in a very similar way to finite groups \cite{Ram_thesis}. 
In fact, we will find that almost all formulae are very similar to 
the formulae used in the symmetric group. 
In terms of the dual basis,
we will obtain a handy expression for 
quantities with upper indices,
\begin{eqnarray}
&&
g^{ij}=
tr^{(r)}(b^ib^j)
\nonumber \\
&&
C_{ij}{}^k=
tr^{(r)}(b_ib_jb^k). 
\end{eqnarray}
In what follows we will exploit the second dual basis $b^i$. 

The computation of 
the two-point function is the same as (\ref{prescription_eta_symmetric}).
We obtain 
\begin{eqnarray}
&&
\eta_{ij}
=\sum_{\gamma}\chi^{\gamma}(b_i)\chi^{\gamma}(b_j)
\nonumber \\
&&
\eta^{ij}
=\sum_{\gamma}\chi^{\gamma}(b^i)\chi^{\gamma}(b^j)
\nonumber \\
&&
\eta_{i}{}^{j}
=\sum_{\gamma}\chi^{\gamma}(b_i)\chi^{\gamma}(b^j).  
\label{projector-brauerTFT}
\end{eqnarray}
If we use  
(\ref{twocharacters_combine_dual}),
we find another expression of the two-point function,
\begin{eqnarray}
&&\eta_{ij}=\sum_{k}tr^{(r)}(b_ib_k b_j b^k)
\nonumber \\
&&\eta^{ij}=\sum_{k}tr^{(r)}(b^ib_k b^j b^k)
\nonumber \\
&&\eta_{i}{}^{j}=\sum_{k}tr^{(r)}(b_ib_k b^j b^k).
\label{projector-brauerTFT_secondform}
\end{eqnarray}
Because $\chi^{\gamma}(b_i)$ is not proportional to $\chi^{\gamma}(b^i)$,
the difference between in-boundaries (corresponding to lower indices)
and out-boundaries (corresponding to upper indices)
is more important.

$\sum_{\gamma}$ denotes the sum over all irreducible representations of the walled 
Brauer algebra.  
An irreducible representation of $B_N(m,n)$ is labelled by a bi-partition 
$(\gamma_+,\gamma_-)$, where $\gamma_+\vdash (m-k)$ and $\gamma_-\vdash (n-k)$, 
and $k$ is an integer in the range $0 \le k \le min(m, n)$.
The sum will be performed by 
\begin{eqnarray}
\sum_{\gamma}=\sum_{k=0}^{min(m,n)}\sum_{\gamma_+\vdash (m-k)}
\sum_{\gamma_-\vdash (n-k)}.
\label{Brauer_Youngdiagram_sum}
\end{eqnarray}

The Hilbert space of physical states is determined 
by the projection operator 
$\eta$. It is the projection onto the centre of the algebra,
\begin{eqnarray}
\eta_{i}{}^{j}b_j
=
\sum_{\gamma}
\chi^{\gamma}(b_i)
\chi^{\gamma}(b^j)b_j
=
\sum_{\gamma}\frac{1}{d_{\gamma}}
\chi^{\gamma}(b_i)
P^{\gamma}.
\label{projection_to_center}
\end{eqnarray}
where $P^{\gamma}$, 
which is given in  
(\ref{projector_central_Brauer}), 
is a central element in the Brauer algebra.
If we define 
\begin{eqnarray}
[[b]]
=
\sum_{i}b_i bb^i, 
\end{eqnarray}
then we can show that 
\begin{eqnarray}
\eta_{i}{}^{j}b_j=[[b_i]].
\end{eqnarray}
Equivalently we have \begin{eqnarray}
\eta_{i}{}^{j}[[b_j]]=[[b_i]].
\end{eqnarray}
Physical operators are invariant under 
time evolution.  

The three-point function is determined by 
the operator product of physical states.
From 
\begin{eqnarray}
[[b_i]][[b_j]]=N_{ij}{}^k[[b_k]],
\end{eqnarray}
we obtain 
\begin{eqnarray}
N_{ijk}=N_{ij}{}^{l}\theta_{lk}
=tr^{(r)}([[b_i]][[b_j]][[b_k]])
=
\sum_{\gamma}
\frac{1}{d^{\gamma}}
\chi^{\gamma}(b_i)\chi^{\gamma}(b_j)\chi^{\gamma}(b_k). 
\label{Brauer_3pt_sphere}
\end{eqnarray}
We can also use 
(\ref{physical_two-point_three-point}) to derive this. 
The associativity 
$([[b_i]][[b_j]])[[b_k]]=[[b_i]]([[b_j]][[b_k]])$ 
is expressed by $N_{ij}{}^kN_{kl}{}^n=N_{ik}{}^n N_{jl}{}^k$. 
The diagonal operator product is realised by 
switching to the representation basis $P^{\gamma}$, that is,  
$P^{\gamma}P^{\gamma^{\prime}}=\delta^{\gamma\gamma^{\prime}}P^{\gamma}$. 
The algebra of $[[b_i]]$ with the bilinear form $\eta$ 
is a commutative Frobenius algebra.

The partition function of a Riemann surface of genus $G$ is  
\begin{eqnarray}
Z_{G}
=\sum_{\gamma}\frac{1}{(d^{\gamma}){}^{2G-2}}, 
\end{eqnarray}
and this is equivalent to  
\begin{eqnarray}
Z_{G}
=
\sum_{i,j}
tr^{(r)}\left(
(c_{i_1}d_{j_1}c^{i_1}d^{j_1})
\cdots 
(c_{i_G}d_{j_G}c^{i_G}d^{j_G})
\right),
\end{eqnarray}
where 
$c,d$ are elements of the Brauer algebra. 
The partition function of $G=1$ gives the dimension of the vector space
\begin{eqnarray}
Z_{G=1}=\sum_{i}\eta_i{}^i
=\sum_{\gamma}1. 
\label{torus_eta_Brauer}
\end{eqnarray}
This counts the number of all irreducible representations.

Likewise for 
a Riemann manifold of genus $G$ with $B_1$ in-boundaries 
and $B_2$ out-boundaries the correlation function is  
\begin{eqnarray}
Z_{G,B_1,B_2}
=\sum_{\gamma}\frac{1}{(d^{\gamma}){}^{2G+B_1+B_2-2}}
\chi^{\gamma}(b_{i_1})\cdots
\chi^{\gamma}(b_{i_{B_1}})
\chi^{\gamma}(b^{j_1})\cdots
\chi^{\gamma}(b^{j_{B_2}}).
\end{eqnarray}
An equivalent form is 
\begin{eqnarray}
&&
Z_{G,B_1,B_2}
=
\sum_{k,l}
tr^{(r)}\left(
(c_{k_1}d_{l_1}c^{k_1}d^{l_1})
\cdots 
(c_{k_G}d_{l_G}c^{k_G}d^{l_G}) 
[[b_{i_1}]] 
\cdots 
[[b_{i_{B_1}}]] 
[[b^{j_1}]]
\cdots 
[[b^{j_{B_2}}]]
\right).
\end{eqnarray}
Both $\sum_{k,l}(c_{k}d_{l}c^{k}d^{l})$ and 
$[[b]]$ are central elements of the Brauer algebra. 
The indices  
are raised or lowered by $\eta$. 
For example 
we can use $\eta^{ij}$ to convert an in-boundary into an out-boundary, finding from
\begin{eqnarray}
\eta^{ij}\chi^{\gamma}(b_j)=\chi^{\gamma}(b^i). 
\end{eqnarray}

There is a formal correspondence between 
the symmetric group $S_{m+n}$ and the walled Brauer algebra $B_N(m,n)$, 
clarified 
thanks to the introduction of the dual basis. 
The correspondence is as follows
\begin{eqnarray}
&&
(R,A) \leftrightarrow (\gamma,A),
\nonumber \\
&& 
t_R\leftrightarrow t_{\gamma},
\nonumber \\
&& 
d_R\leftrightarrow d_{\gamma},
\nonumber \\
&& tr^{(r)}(\sigma) \leftrightarrow tr^{(r)}(b), 
\nonumber \\
&& \sigma_i\leftrightarrow b_i, 
\nonumber \\
&& \sigma^i\leftrightarrow b^i. 
\label{formal_replacement}
\end{eqnarray}
The only difference between them is that in the symmetric group 
$\sigma_i\sigma^i$ is proportional to the unit element, while in the Brauer algebra 
$b_i b^i$ is not proportional to the unit, where 
the repeated index $i$ is not summed in each case.


\subsection{Brauer topological field theory as a collection of 
symmetric group topological field theories}
\label{BrauerTFT_by_symmetricTFT}

The partition functions 
obtained from the symmetric group 
in section \ref{sec:TFT_review}
can be regarded as counting of $n$-fold coverings 
of a general Riemann manifold, and this idea plays 
a crucial role in the string theoretic description of 
large $N$
two-dimensional Yang-Mills theory \cite{GrossTaylor,9402107,9411210}.
On the other hand, 
an interpretation in terms of coverings 
is less clear for 
walled Brauer algebras, 
but it is naturally expected to give two kinds of coverings because 
the walled Brauer algebra contains the group algebra of $S_m\times S_n$ 
as a subalgebra. 
In fact, in \cite{GrossTaylor}
the coupled representations, 
which are irreducible representations of the Brauer algebra corresponding to $k=0$, 
were considered to describe 
non-holomorphic maps from worldsheets to the target space.
In \cite{0802.3662} the full expansion of two-dimensional Yang-Mills 
is described in terms of only holomorphic maps 
from 
a new formula for the couple representation of $GL(N)$ derived 
with the help of the walled Brauer algebra. 

In this subsection, we will show that the correlation functions 
of the Brauer topological field theory 
can be expressed in terms of 
correlation functions obtained from 
the symmetric group $S_{m-k}\times S_{n-k}$, where 
$k$ takes all integers in $1\le k\le min(m,n)$.

The partition functions of a surface of genus $G$
are easily expressed in terms of the symmetric group data, 
if we use the formula (\ref{Brauer_dimension_symmetric}), as 
\begin{eqnarray}
Z_{G}^{B_N(m,n)}
&=&\sum_{\gamma}\frac{1}{(d^{\gamma}){}^{2G-2}}
\nonumber \\
&=&
\sum_{k=0}^{min(m,n)}
\sum_{\gamma_+\vdash (m-k)}\sum_{\gamma_-\vdash (n-k)}
\left(
\frac{(m-k)!(n-k)!k!}{m!n!}
\right)^{2-2G}
\frac{1}{(d^{\gamma_+}){}^{2G-2}}\frac{1}{(d^{\gamma_-}){}^{2G-2}}
\nonumber \\
&=&
\sum_{k=0}^{min(m,n)}
\left(
\frac{(m-k)!(n-k)!k!}{m!n!}
\right)^{2-2G}
Z_{G}^{S_{m-k}}Z_{G}^{S_{n-k}}, 
\label{Brauer_intermsof_symmetric_partition}
\end{eqnarray}
where 
$Z_{G}^{S_{m-k}}$ is the partition function (\ref{partition_function_symmetric_G}) 
corresponding to $S_{m-k}$,
\begin{eqnarray}
Z_{G}^{S_{m-k}}=\sum_{R\vdash (m-k)}
\frac{1}{(d_R)^{2G-2}}. 
\end{eqnarray}
Because the right-hand side 
in (\ref{Brauer_intermsof_symmetric_partition}) 
has a clear interpretation in terms of coverings of the Riemann surface of 
genus $G$, 
the partition function of  
the Brauer topological field theory can have a meaning 
in terms of $(m-k)$-coverings and $(n-k)$-coverings 
of the Riemann surface. 
It is interesting that a smaller number of sheets than $m,n$ 
also come in the game.

\begin{figure}[t]
\begin{center}
\includegraphics[scale=0.5]{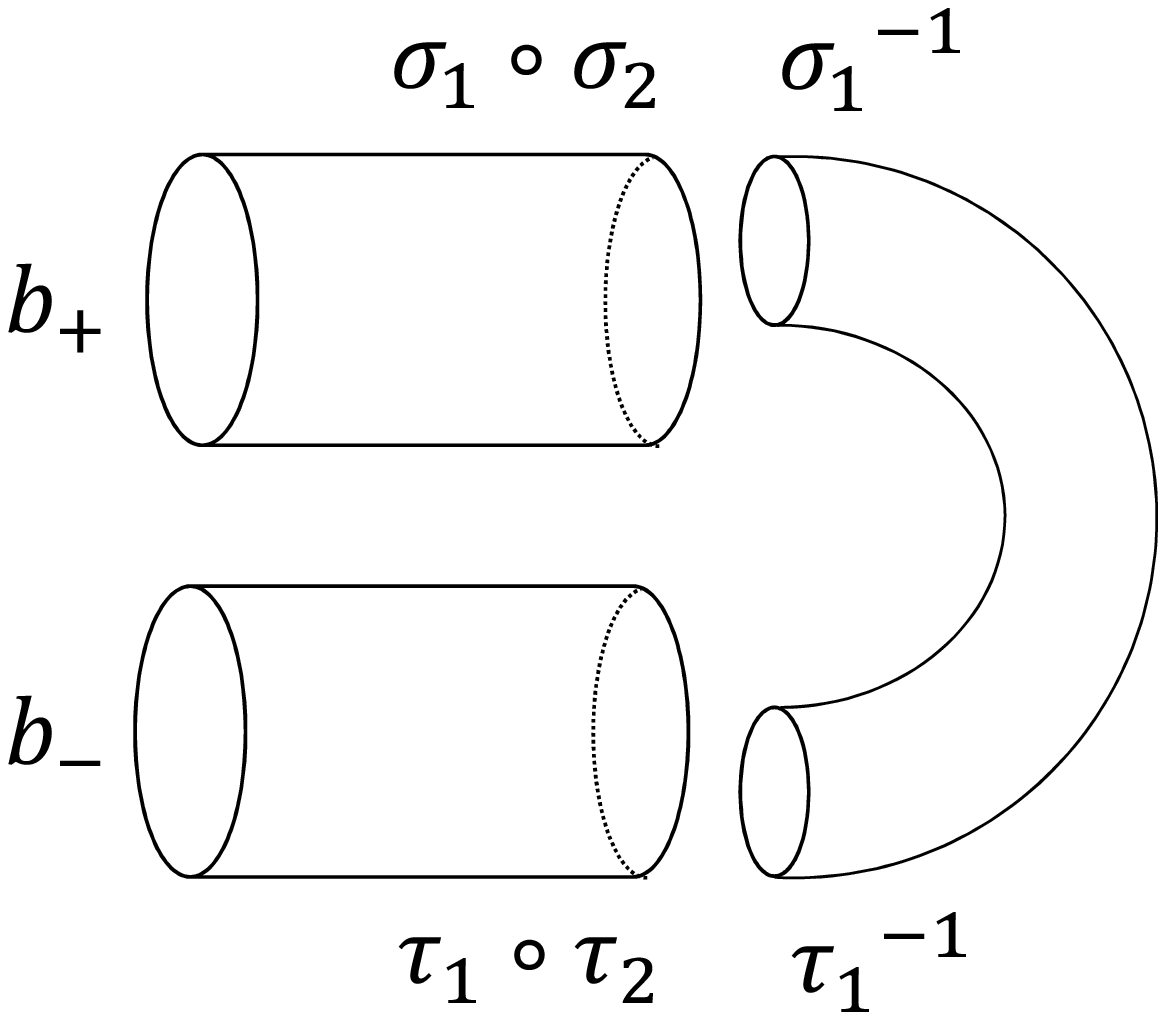}
\caption{
pictorial drawing of (\ref{Delta_cylinders})
} 
\label{fig:Delta_three_cylinders}
\end{center}
\end{figure}

We next consider correlation functions. 
For the purpose 
we will use 
a character formula of the walled Brauer algebra.
The character of a general element 
$b$ in the walled Brauer algebra 
can be expressed in terms of the character of an element of the form 
$C^{\otimes h}\otimes b_+\otimes b_-$ as 
\begin{eqnarray}
\chi^{\gamma}(b)=N^{z(b)-h}\chi^{\gamma}(C^{\otimes h}\otimes b_{+}\otimes b_{-}), 
\label{character_formula_brauer_1}
\end{eqnarray}
where $b_+$ and $b_-$ are elements in $S_{m-h}$ and $S_{n-h}$ respectively,  
and $C$ is the contraction.
The $z(b)$ is a quantity that is read from the element $b$. 
This formula is explained in  
(\ref{Brauer_character_general}). 
Then the two-point function can be   
\begin{eqnarray}
\eta^{B_N(m,n)}(b,\tilde{b})
&=&\sum_{\gamma}\chi^{\gamma}(b)\chi^{\gamma}(\tilde{b})
\nonumber \\
&=&
N^{z(b)+z(\tilde{b})-h-\tilde{h}}
\eta^{B_N(m,n)}(C^{\otimes h}\otimes b_{+}\otimes b_{-},
C^{\otimes \tilde{h}}\otimes \tilde{b}_{+}\otimes \tilde{b}_{-}).
\end{eqnarray}
We will next use 
\begin{eqnarray}
\chi^{\gamma}(C^{\otimes h}\otimes b_{+}\otimes b_{-})
=
\frac{N^h }{(m-k)!(n-k)!}
\sum_{\sigma_2 \in S_{m-k},\tau_2\in S_{n-k}}
\Delta(b_+,b_-;\sigma_2,\tau_2)\chi_{\gamma_+}(\sigma_2^{-1})
\chi_{\gamma_-}(\tau_2^{-1}),
\label{formula_new_Brauer_character}
\end{eqnarray}
where 
\begin{eqnarray}
\Delta(b_+,b_-;\sigma_2,\tau_2)
=\frac{1}{(k-h)!^2}
\sum_{\sigma_1,\tau_1\in S_{k-h}}
\eta^{S_{m-h}}(b_+,\sigma_1\circ \sigma_2)
\eta^{S_{n-h}}(b_-,\tau_1\circ \tau_2)
\eta^{S_{k-h}}(\sigma_1^{-1}, \tau_1^{-1}). 
\label{Delta_cylinders}
\end{eqnarray}
This is derived in (\ref{brauer_character_symmetric_group}). 
The $\Delta$ is composed of three two-point functions defined in 
the topological field theory of the symmetric group
(see figure \ref{fig:Delta_three_cylinders}).

The formula enables us to compute 
\begin{eqnarray}
&&
\eta^{B_N(m,n)}(b,\tilde{b})
\nonumber \\ 
&=&
N^{z(b)+z(\tilde{b})-h-\tilde{h}}
\eta^{B_N(m,n)}(C^{\otimes h}\otimes b_{+}\otimes b_{-},
C^{\otimes \tilde{h}}\otimes \tilde{b}_{+}\otimes \tilde{b}_{-})
\nonumber \\ 
&=&
N^{z(b)+z(\tilde{b})-h-\tilde{h}}
\sum_{\gamma}\chi^{\gamma}(C^{\otimes h}\otimes b_{+}\otimes b_{-})
\chi^{\gamma}(C^{\otimes \tilde{h}}\otimes \tilde{b}_{+}\otimes \tilde{b}_{-})
\nonumber \\
&=&
N^{z(b)+z(\tilde{b})}
\sum_{k=max(h,\tilde{h})}^{min(m,n)}
\sum_{\gamma_+ \vdash (m-k)}
\sum_{\gamma_- \vdash (n-k)}
\nonumber \\
&&
\times \frac{1}{(m-k)!(n-k)!}
\sum_{\sigma_2 \in S_{m-k}}
\sum_{\tau_2 \in S_{n-k}}
\Delta(b_+,b_-;\sigma_2,\tau_2)
\chi_{\gamma_+}(\sigma_2^{-1})
\chi_{\gamma_-}(\tau_2^{-1})
\nonumber \\
&&
\times \frac{1}{(m-k)!(n-k)!}
\sum_{\tilde{\sigma}_2 \in S_{m-k}}
\sum_{\tilde{\tau}_2 \in S_{n-k}}
\Delta(\tilde{b}_+,\tilde{b}_-;\tilde{\sigma}_2,\tilde{\tau}_2)
\chi_{\gamma_+}(\tilde{\sigma}_2^{-1})
\chi_{\gamma_-}(\tilde{\tau}_2^{-1})
\nonumber \\ 
&=&
N^{z(b)+z(\tilde{b})}
\sum_{k=max(h,\tilde{h})}^{min(m,n)}
\sum_{\sigma_2,\tilde{\sigma}_2\in S_{m-k}}
\sum_{\tau_2,\tilde{\tau}_2\in S_{n-k}}
\Delta(b_+,b_-;\sigma_2,\tau_2)
\Delta(\tilde{b}_+,\tilde{b}_-;\tilde{\sigma}_2,\tilde{\tau}_2)
\nonumber \\
&&
\times 
\eta^{S_{m-k}\times S_{n-k}}
(\sigma_2^{-1}\otimes \tau_2^{-1},\tilde{\sigma}_2^{-1}\otimes  \tilde{\tau}_2^{-1}),
\label{correlator_brauer_by_symmetric}
\end{eqnarray}
where $\eta$ is the two-point function (\ref{projector-symmetric}) 
corresponding to $S_{m-k}\times S_{n-k}$,
\begin{eqnarray}
\eta^{S_{m-k}\times S_{n-k}}
(\sigma_2^{-1}\otimes \tau_2^{-1},\tilde{\sigma}_2^{-1}\otimes  \tilde{\tau}_2^{-1})
=
\eta^{S_{m-k}}
(\sigma_2^{-1},\tilde{\sigma}_2^{-1})
\eta^{S_{n-k}}
(\tau_2^{-1},\tilde{\tau}_2^{-1}),
\end{eqnarray}
and 
\begin{eqnarray}
\eta^{S_{m-k}}
(\sigma_2^{-1},\tilde{\sigma}_2^{-1})
=\frac{1}{((m-k)!)^2}\sum_{\gamma_+\vdash (m-k)}
\chi_{\gamma_+}(\sigma_2^{-1})\chi_{\gamma_+}(\tilde{\sigma}_2^{-1}).
\end{eqnarray}
We have shown that 
the two-point function of the Brauer topological field theory
can be expressed in terms of 
the two-point function of the topological field theory obtained from 
the group algebra of 
$S_{m-k}\times S_{n-k}$, where $k$ takes all integers in 
$max(h,\tilde{h})\le k\le min(m,n)$. 

It is straightforward to extend this to any correlation functions. 
For example 
the three-holed sphere (\ref{Brauer_3pt_sphere}) is as follows:
\begin{eqnarray}
&&
Z_{G=0,B=3}^{B_N(m,n)}(b^1,b^2,b^3)
\nonumber \\
&=&
N^{z(b^1)+z(b^2)+z(b^3)-h_1-h_2-h_3}
Z_{G=0,B=3}^{B_N(m,n)}(C^{\otimes h_1}\otimes b_{+}^1\otimes b_{-}^1,
C^{\otimes h_2}\otimes b_{+}^2\otimes b_{-}^2,
C^{\otimes h_3}\otimes b_{+}^3\otimes b_{-}^3,)
\nonumber \\
&=&
N^{z(b^1)+z(b^2)+z(b^3)}
\sum_{k=max(h_1,h_2,h_3)}^{min(m,n)}
\sum_{\sigma_1,\sigma_2,\sigma_3\in S_{m-k}}
\sum_{\tau_1,\tau_2,\tau_3\in S_{n-k}}
\nonumber \\
&&
\times 
\Delta(b_+^1,b_-^1;\sigma_1,\tau_1)
\Delta(b_+^2,b_-^2;\sigma_2,\tau_2)
\Delta(b_+^3,b_-^3;\sigma_3,\tau_3)
\nonumber \\
&&
\times \frac{k!(m-k)!(n-k)!}{m!n!}
Z_{G=0,B=3}^{S_{m-k}\times S_{n-k}}
(\sigma_1^{-1}\otimes \tau_1^{-1},
\sigma_2^{-1}\otimes \tau_2^{-1},
\sigma_3^{-1}\otimes \tau_3^{-1}),
\end{eqnarray}
where 
$Z_{G=0,B=3}^{S_{m-k}\times S_{n-k}}
(\sigma_1^{-1}\otimes \tau_1^{-1},
\sigma_2^{-1}\otimes \tau_2^{-1},
\sigma_3^{-1}\otimes \tau_3^{-1})$ is the three-holed sphere
given in (\ref{three_point_etaTFT}) 
associated with 
$S_{m-k}\times S_{n-k}$.
Thus we can express any correlation functions 
obtained from the Brauer algebra as a set of
correlation functions obtained from the symmetric group 
$S_{m-k}\times S_{n-k}$. 


\section{2D field theories with the restricted structure}
\label{sec:new_theory_subgroup}

As we have reviewed in section \ref{sec:N=4_review}, 
gauge invariant operators in ${\cal N}=4$ super Yang-Mills 
built from $p$ complex matrices 
can be labelled by an element of the form 
$[\sigma_i]_H$ or $[b_i]_H$, where 
$H=S_{n_1}\times \cdots \times  S_{n_p}$.  
With the motivation to associate 
the description of the matrix models with two-dimensional field theories, 
we will construct a class of two-dimensional 
field theories whose physical states are given by 
$[\sigma_i]_H$ or $[b_i]_H$. 
The idea of such theories is given in \cite{1301.1980}. 
For simplicity 
in what follows we will consider the case
$H=S_{m}\times S_n$, but the generalisation to $S_{n_1}\times \cdots \times S_{n_p}$ 
is straightforward.

\begin{figure}[t]
\begin{center}
 \includegraphics[scale=0.6]{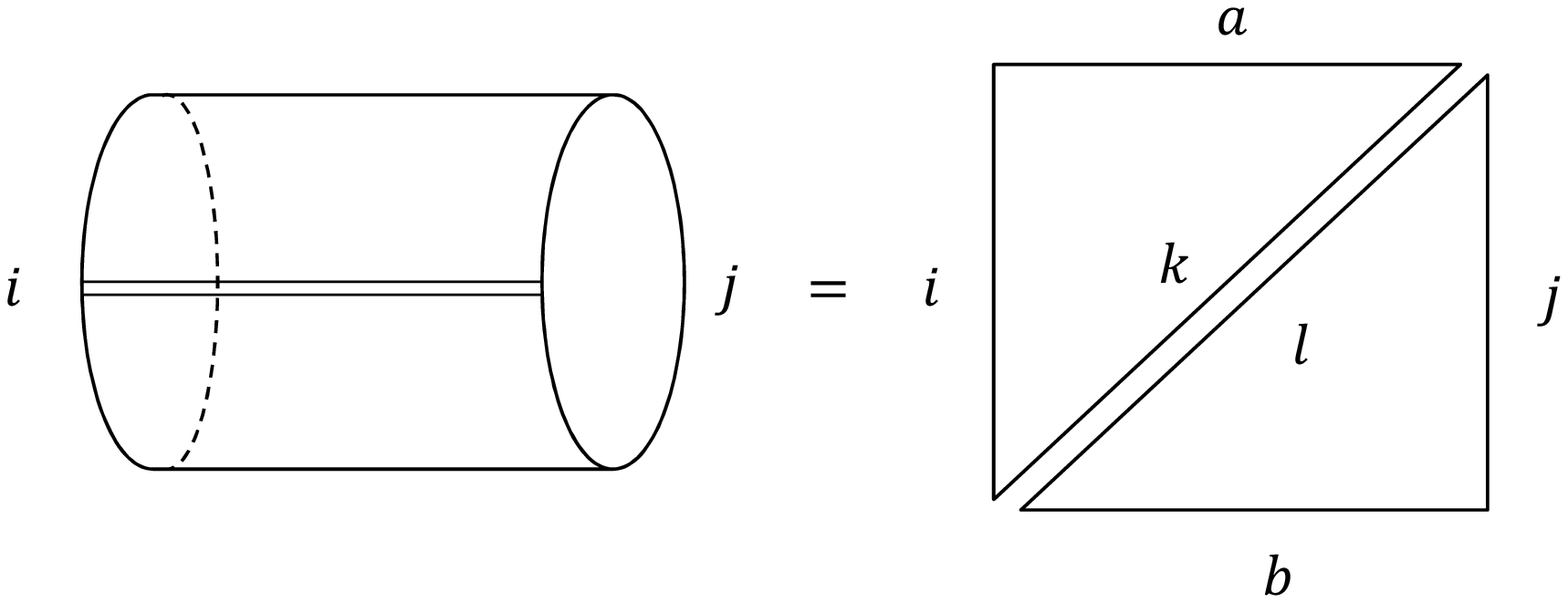}
\caption{
LHS - 
a cylinder with the restriction on the double line, 
RHS - a simple triangulation 
} 
\label{fig:cylinderH}
\end{center}
\end{figure}

We will first consider the symmetric group $S_{m+n}$. 
As we have seen in the last two sections, 
physical operators are determined by 
the two-point function, 
because it is a projection operator onto the Hilbert space 
of physical operators.
In order to obtain physical operators labelled by $[\sigma_i]_H$, 
we will consider the two-point function
drawn in figure \ref{fig:cylinderH} 
in stead of the cylinder (\ref{prescription_eta_symmetric}). 
On the double line in the figure we restrict the sum to the subgroup 
$H$ \cite{1301.1980}, and  
triangulations should be done consistently with the restriction. 
A simple triangulation of such a cylinder is shown in the RHS of 
figure \ref{fig:cylinderH}:
\begin{eqnarray}
\theta_{ij}=\sum_{k}\sum_a C_{ikb}C_{ajl} h^{ab}g^{kl},
\label{theta_CC}
\end{eqnarray}
where $a$ runs over 
a complete set of the subgroup $H$.
The $g^{ij}$ and $C_{ijk}$ are 
defined in (\ref{metric_C_symmetric_def}) and (\ref{inverse_g^ij}).
The $h$ is the metric defined in $H$ as 
\begin{eqnarray}
&&
h_{ab}=m!n!\delta_{m,n}(\sigma_a\sigma_b)
\nonumber \\
&&
h^{ab}=m!n!\delta_{m,n}(\sigma^a\sigma^b),
\label{metric_subalgebra}
\end{eqnarray}
where $\sigma^a$ is the dual basis of $\sigma_a$, which is defined by 
$\sigma^a=\frac{1}{m!n!}\sigma_a^{-1}$. 
The delta function defiend over  
the group algebra of $H$ is denoted by  
$\delta_{m,n}(\sigma)$. 

We can show that 
\begin{eqnarray}
&&
\theta_{ij}
=\sum_{a}tr^{(r)}(h_a \sigma_ih^a\sigma_j)
=tr^{(r)}([\sigma_i]_H[\sigma_j]_H)
\nonumber \\
&&
\theta^{ij}
=\sum_{a}tr^{(r)}(h_a\sigma^ih^a\sigma^j)
=tr^{(r)}([\sigma^i]_H[\sigma^j]_H)
\nonumber \\
&&
\theta_{i}{}^{j}=\sum_{a}tr^{(r)}(h_a\sigma_ih^a\sigma^j)
=tr^{(r)}([\sigma_i]_H[\sigma^j]_H),
\label{theta_regularrep}
\end{eqnarray}
and the projector relation $\theta^2=\theta$
can be checked explicitly.
The use of (\ref{Qoperator_characer_combine_symmetric}) leads to another 
expression: 
\begin{eqnarray}
&&
\theta_{ij}
=\sum_{R,A,\mu,\nu}
\frac{d^{R}}{d_A}
\chi^{R}_{A,\mu\nu}(\sigma_i)\chi^{R}_{A,\nu\mu}(\sigma_j)
\nonumber \\
&&
\theta^{ij}=
\sum_{R,A,\mu,\nu}
\frac{d^{R}}{d_A}
\chi^{R}_{A,\mu\nu}(\sigma^i)
\chi^{R}_{A,\nu\mu}(\sigma^j)
\nonumber \\
&&
\theta_{i}{}^{j}=
\sum_{R,A,\mu,\nu}
\frac{d^{R}}{d_A}
\chi^{R}_{A,\mu\nu}(\sigma_i)\chi^{R}_{A,\nu\mu}(\sigma^j),
\label{theta_restricted_character_symmetric} 
\end{eqnarray}
where $d_R$ is the dimension of 
an irreducible representation $R$ of the symmetric group $S_{m+n}$, while 
 $d_A$ is the dimension of 
an irreducible representation $A$ of the symmetric group $S_{m}\times S_n$. 
The $\chi^{R}_{A,\mu\nu}(\sigma)$ is the restricted character 
(see around (\ref{intertwiner_symmetric_group})). 
The indices $\mu,\nu$ run over 
$1,\cdots,M^R_A$, where $M^R_A$ 
counts the number of times the representation $A$ appears in
the $R$.
This projector 
(\ref{theta_restricted_character_symmetric}) 
was obtained in \cite{1301.1980} 
from 
the invariance (\ref{equivalence_class_two_matrix_symmetric}), 
which leads to 
$\theta_{i}{}^{j}
tr_{m+n}(\sigma_j X^{\otimes m}\otimes Y^{\otimes n})
=tr_{m+n}(\sigma_i X^{\otimes m}\otimes Y^{\otimes n})$. 
We have another expression \cite{1301.1980}
which is closely related to \cite{0711.0176}, 
instead of 
(\ref{theta_restricted_character_symmetric}).  

From the cyclic property of the restricted character 
\begin{eqnarray}
&&
\chi^R_{A,\mu\nu}(h\sigma)=\chi^R_{A,\mu\nu}(\sigma h)
\quad (h\in S_m\times S_n)
\nonumber \\
&&
\chi^R_{A,\mu\nu}(\rho \sigma)\neq \chi^R_{A,\mu\nu}(\sigma \rho)
\quad (\rho \in  S_{m+n}/S_m\times S_n), 
\end{eqnarray}
boundary operators are more sensitive to the orientations corresponding to 
upper and lower indices 
because 
$\sigma_i$ is not conjugate by 
some permutation in the subgroup $S_m\times S_n$ to 
the dual element $\sigma^i$.

Physical operators are determined by the projection operator
\begin{eqnarray}
\theta_{i}{}^{j}\sigma_j
=
\sum_{R,A,\mu,\nu}\frac{1}{d_R}
\chi^{R}_{A,\mu\nu}(\sigma_i)
P^{R}_{A,\mu\nu}=[\sigma_i]_H, 
\label{restriction_another_projector}
\end{eqnarray}
where $[\sigma_i]_H$ is defined in 
(\ref{conjugacy_class_subgroup_symbol}), 
and $P^{R}_{A,\mu\nu}$ is a basis of elements commuting with 
any elements 
in $H=S_{m}\times S_n$ \cite{0801.2061}. 
See (\ref{intertwiner_symmetric_group}).
The equation 
(\ref{restriction_another_projector}) also means that physical states are invariant under time evolution, 
\begin{eqnarray}
\theta_{i}{}^{j}[\sigma_j]_H=[\sigma_i]_H. 
\end{eqnarray}
Because $\theta_i{}^j$ is a projector, 
we expect to obtain the 
the dimension of the vector space 
by taking a trace in the same way as the topological field theories in section 
\ref{sec:TFT_review} and \ref{topological_Brauer}, 
\begin{eqnarray}
Z_{G=1}=
\sum_{i}
\theta_{i}{}^{i}
=\sum_{R,A}(M^{R}_A)^2,
\label{counting_new_2D_symmetric}
\end{eqnarray}
which comes from (\ref{orthogonality_restricted_character_symmetric}). 
This is the partition function of a torus. 
The dimension of the vector space gives the expected result of the counting of 
the number of 
multi-traces constructed from $m$ copies of $X$ and $n$ copies of $Y$ via the method of 
using the basis (\ref{two-matrix_symmetric}), 
as shown in \cite{0801.2061,0810.4217}.
The counting of multi-traces is 
associated with $Z_{G=1}$ by considering the counting of 
orbits of the group actions via Burnside's lemma in
\cite{1301.1980,1110.4858}. 

The three-point function (three-holed sphere) $\Xi_{ijk}$
can be obtained as the structure constant of 
the algebra spanned by $[\sigma_i]_H$. 
From  
\begin{eqnarray}
[\sigma_i]_H[\sigma_j]_H=\Xi_{ij}{}^k[\sigma_k]_H, 
\end{eqnarray}
we find that 
\begin{eqnarray}
&&
\Xi_{ij}{}^{k}
=tr^{(r)}([\sigma_i]_H[\sigma_j]_H[\sigma^k]_H),
\nonumber \\
&&
\Xi_{ijk}
=tr^{(r)}([\sigma_i]_H[\sigma_j]_H[\sigma_k]_H)=\Xi_{ij}{}^{l}\theta_{lk}.
\label{3pt_thetaTFT_symmetric}
\end{eqnarray}
Note that 
the two-point function and the three-point function 
are also obtained by acting with the projection operator 
on $g_{ij}$ and $C_{ijk}$  
\begin{eqnarray}
&&
\theta_{i}{}^{k}\theta_{j}{}^{l}g_{kl}=
\theta_{ij}
\nonumber \\
&&
\theta_{i}{}^{p}\theta_{j}{}^{q}\theta_{k}{}^{r} 
C_{pqr}=\Xi_{ijk},
\end{eqnarray}
which is analogous to (\ref{physical_two-point_three-point}). 
It is also convenient to rewrite the expression in terms of the restricted characters:
\begin{eqnarray}
\Xi_{ij}{}^{k}
=\sum_{R,A,\mu,\nu,\lambda}
\frac{d^{R}}{(d_A)^2}
\chi^{R}_{A,\mu\nu}(\sigma_i)
\chi^{R}_{A,\nu\lambda}(\sigma_j)
\chi^{R}_{A,\lambda\mu}(\sigma^k).
\label{three-point_function}
\end{eqnarray}
The associativity of the algebra, 
$([\sigma_i]_H[\sigma_j]_H)[\sigma_k]_H=[\sigma_i]_H([\sigma_j]_H[\sigma_k]_H)$, 
is translated into 
\begin{eqnarray}
\Xi_{ij}{}^k\Xi_{kl}{}^n=\Xi_{ik}{}^n\Xi_{jl}{}^k. 
\end{eqnarray}

The vector space whose elements are $[\sigma]_H$, equipped with 
the bilinear form $\theta_{ij}$,  
is a Frobenius algebra.  
The Frobenius form is  $\theta_{i0}$. 
One can check 
\begin{eqnarray}
\Xi_{ij}{}^k\theta_{k0}=\theta_{ij}.
\end{eqnarray}
A big difference from the Frobenius algebras 
in section \ref{sec:TFT_review} and \ref{topological_Brauer}
is that the Frobenius algebra in this section
is {\it noncommutative}, that is, 
\begin{eqnarray}
\Xi_{ij}{}^k\neq \Xi_{ji}{}^k 
\label{NC_Frobenius_three_point}
\end{eqnarray}
as a consequence of the non-commutative operator algebra 
\begin{eqnarray}
[\sigma_i]_H[\sigma_j]_H\neq [\sigma_j]_H[\sigma_i]_H. 
\end{eqnarray}
By a change of basis, we will obtain the following (almost diagonal) 
operator product
\begin{eqnarray}
P^{R}_{A,\mu\nu}
P^{R^{\prime}}_{A^{\prime},\mu^{\prime}\nu^{\prime}}
=\delta^{R R^{\prime}}
\delta_{A A^{\prime}}
\delta_{\nu \mu^{\prime}}
P^{R}_{A,\mu \nu^{\prime}}. 
\end{eqnarray}
The existence of the $\mu,\nu$ indices reflects the noncommutativity, and 
we also find 
the $\mu,\nu$ indices indeed behave like matrix-indices. 
(See also orthogonality relations for restricted characters 
in appendix \ref{sec:formula_orthogonality}.)
Due to the non-commutativity, this field theory is not topological.

The partition functions can be computed 
from the building blocks $\theta_{ij},\Xi_{ijk}$. 
The partition function of a Riemann surface of genus $G$
can be computed as
\begin{eqnarray}
\sum_{R,A}
\frac{1}{(d^R d_A)^{G-1}}(M^R_A)^{G+1}. 
\label{genusG_restricted}
\end{eqnarray}
Equations in 
appendix \ref{sec:formula_orthogonality}
will be helpful to derive this. 
The partition function depends only on $G$. 
If we set $M^R_A=1$ and $d_A=d_R$, we reproduce the 
result (\ref{partition_function_symmetric_G}) as expected. 
The partition function has another form,
\begin{eqnarray}
Z_{G}
=
\sum_{a,j}
tr^{(r)}\left(
(h_{a_1}\sigma_{j_1}h^{a_1}\sigma^{j_1})
\cdots 
(h_{a_G}\sigma_{j_G}h^{a_G}\sigma^{j_G})
\right),
\end{eqnarray}
where $h_a$ is a basis of the subgroup $S_m\times S_n$ and 
$h^a$ is the dual basis. 
The indices $a_1,\cdots ,a_G$ run over a complete set of $S_m\times S_n$.   

The next interest will be a Riemann surface with boundaries. 
This should be paid more 
attention because the Frobenius algebra is noncommutative.
Correlation functions depend on 
the way of gluing the building blocks.
As a simple example consider the case of $G=1$ and $B=2$. 
We have two possible ways,  
$\Xi_{ijk}\Xi^{jkl}$ and $\Xi_{ijk}\Xi^{kjl}$, giving different answers. 
The first one is computed to give 
\begin{eqnarray}
\Xi_{ijk}\Xi^{jkl}
=\sum_{R,A,\mu,\nu}
\frac{1}{(d_A)^{2}}
\chi^{R}_{A,\mu\mu}(\sigma_{i})
\chi^{R}_{A,\nu\nu }(\sigma^{l}), 
\end{eqnarray}
while the second one is 
\begin{eqnarray}
\Xi_{ijk}\Xi^{kjl}
=
\sum_{R,A,\mu,\nu}
\frac{1}{(d_A)^{2}} M^R_A
\chi^{R}_{A,\mu\nu}(\sigma_{i})
\chi^{R}_{A,\nu\mu }(\sigma^{l}). 
\end{eqnarray}
The second one is the same as $\Xi_{i}{}^l{}_{k}\Xi_{p}{}^{pk}$. 
If we set $\sigma_i=1$ and $\sigma^l=1$, both reduce to the case 
$G=1$ in (\ref{genusG_restricted}). 
The difference between the two 
correlation functions 
is how multiplicity indices $\mu,\nu$ are contracted. 
Introducing   
$M^R_A\times M^R_A$ matrices $M_i$,  
the structure of $tr(M_1)tr(M_2)$ arises effectively from the first one, 
while $tr1tr(M_1M_2)$ from the second one.
The first one can also be written as 
\begin{eqnarray}
\Xi_{ijk}\Xi^{jkl}
=
\sum_{a,p}
tr^{(r)}(h_{a} \sigma_{p} [\sigma_i]_H h^{a}\sigma^{p} [\sigma^l]_H),
\end{eqnarray}
while the second can be 
\begin{eqnarray}
\Xi_{ijk}\Xi^{kjl}
=
\sum_{a,p}
tr^{(r)}(h_{a}\sigma_{p}h^{a}\sigma^{p}[\sigma_i]_H [\sigma^l]_H).
\end{eqnarray}
Note that 
\begin{eqnarray}
 [\sigma_i]_H h= h[\sigma_i]_H \quad (h\in S_m\times S_n),
\end{eqnarray}
but 
\begin{eqnarray}
 [\sigma_i]_H \tau \neq \tau[\sigma_i]_H \quad (\tau \in S_{m+n}/S_m\times S_n).
\end{eqnarray}
As we have seen in the example, the correlation functions are not uniquely determined 
when we specify $G$ and $B$, and they are sensitive to the positions of boundaries. 
Reflecting the matrix structure, 
they have cyclic symmetries acting on the boundary positions. 
For example the following correlation function, 
which is one of correlation functions of a surface of genus $G$ with $B$ boundaries, 
\begin{eqnarray}
\sum_{R,A}
\frac{1}{(d^R)^{G-1} (d_A)^{G+B-1}}(M^R_A)^{G}
\sum_{\mu,\nu,\tau}
\chi^{R}_{A,\mu\nu}(\sigma_{i_1})
\chi^{R}_{A,\nu\lambda }(\sigma_{i_2})
\cdots
\chi^{R}_{A,\tau\mu}(\sigma_{i_B})
\end{eqnarray}
is invariant under cyclic permutations acting on the indices $i_1,\cdots,i_B$. 
One may read the single trace structure $tr(M_1\cdots M_B)$ from this. 
The restricted characters were first introduced to describe 
open strings on giant gravitons
\cite{0411205,0701066}. 
It is interesting to study if the origin of of 
the noncommutativity can be explained in terms of 
D-branes in the context of two-dimensional theories.

\qquad

So far 
we have considered a new kind of two-dimensional field theories based on the symmetric group, 
and the same construction can be applied to the walled Brauer algebra. 
The formal replacement (\ref{formal_replacement})
with 
\begin{eqnarray}
&&[\sigma_i]_H \leftrightarrow [b_i]_H,
\nonumber \\
&&
M^R_A \leftrightarrow M^{\gamma}_A ,
\end{eqnarray}
where $M^{\gamma}_A$ is 
the number of times $A$ appears in $\gamma$, 
will work to obtain 
the new two-dimensional 
field theory based on the walled Brauer algebra. 
We will show some of them explicitly for convenience. 
The two-point function is given by 
\begin{eqnarray}
&&
\theta_{ij}
=
tr^{(r)}([b_i]_H[b_j]_H)=
\sum_{\gamma,A,\mu,\nu}
\frac{d^{\gamma}}{d_A}
\chi^{\gamma}_{A,\mu\nu}(b_i)\chi^{\gamma}_{A,\nu\mu}(b_j)
\nonumber \\
&&
\theta^{ij}=tr^{(r)}([b^i]_H[b^j]_H)=
\sum_{\gamma,A,\mu,\nu}
\frac{d^{\gamma}}{d_A}
\chi^{\gamma}_{A,\mu\nu}(b^i)
\chi^{\gamma}_{A,\nu\mu}(b^j)
\nonumber \\
&&
\theta_{i}{}^{j}=tr^{(r)}([b_i]_H[b^j]_H)=
\sum_{\gamma,A,\mu,\nu}
\frac{d^{\gamma}}{d_A}
\chi^{\gamma}_{A,\mu\nu}(b_i)\chi^{\gamma}_{A,\nu\mu}(b^j), 
\label{projector-brauerTFT_subgroup}
\end{eqnarray}
and this determines 
the Hilbert space of physical operators, 
\begin{eqnarray}
\theta_{i}{}^{j}b_j
=
\sum_{\gamma,A,\mu,\nu}\frac{1}{d_A}
\chi^{\gamma}_{A,\mu\nu}(b_i)
Q^{\gamma}_{A,\mu\nu}=[b_i]_H. 
\end{eqnarray}
Changing to the basis (\ref{explicit_form_Q-operator}), 
the almost diagonal operator product is obtained \cite{0709.2158,0807.3696}
\begin{eqnarray}
Q^{\gamma}_{A,\mu\nu}
Q^{\gamma^{\prime}}_{A^{\prime},\mu^{\prime}\nu^{\prime}}
=\delta^{\gamma \gamma^{\prime}}
\delta_{A A^{\prime}}
\delta_{\nu \mu^{\prime}}
Q^{\gamma}_{A,\mu \nu^{\prime}}, 
\end{eqnarray}
where $\mu,\nu$ are multiplicity labels 
running over $1,\cdots,M^{\gamma}_A$.
The three-point function is the structure constant 
(i.e. $[b_i]_H[b_j]_H=\Xi_{ij}{}^{k}[b_k]_H$), 
\begin{eqnarray}
\Xi_{ijk}
=
tr^{(r)}([b_i]_H[b_j]_H[b_k]_H)
=
\sum_{\gamma,A,\mu,\nu,\lambda}
\frac{d^{\gamma}}{(d_A)^2}
\chi^{\gamma}_{A,\mu\nu}(b_i)
\chi^{\gamma}_{A,\nu\lambda}(b_j)
\chi^{\gamma}_{A,\lambda\mu}(b_k).
\end{eqnarray}
The partition function of a Riemann surface of genus $G$ is given by
\begin{eqnarray}
Z_{G}
&=&
\sum_{a,j}
tr^{(r)}\left(
(h_{a_1}b_{j_1}h^{a_1}b^{j_1})
\cdots 
(h_{a_G}b_{j_G}h^{a_G}b^{j_G})
\right)
\nonumber \\
&=&
\sum_{\gamma,A}
\frac{1}{(d^{\gamma} d_A)^{G-1}}(M^{\gamma}_A)^{G+1}.
\end{eqnarray}
Furnished 
with the bilinear form (\ref{projector-brauerTFT_subgroup}), 
the algebra formed by $[b_i]_H$ is a noncommutative Frobenius algebra. 
The partition function of $G=1$ gives the dimension of the vector space 
\begin{eqnarray}
Z_{G=1}=
\sum_i\theta_{i}{}^{i}
=\sum_{\gamma,A}(M^{\gamma}_A)^2. 
\label{counting_new_2D_Brauer}
\end{eqnarray}
This is equivalent to the counting of 
the number of 
multi-traces built from $m$ copies of $X$ and $n$ copies of $Y$ via the method of 
using the basis (\ref{two-matrix_brauer}), as shown in \cite{0709.2158,0911.4408}.


\section{2D theoretic interpretation of the multi-matrix models}
\label{N=4SYM_2pt_TFT}

In this section we will study a relation between 
correlators of the Gaussian matrix models and 
correlation functions 
of the two-dimensional field theories. 

First consider the method of using symmetric group elements 
to describe multi-trace operators. 
Recall (\ref{two-point_two-matrix_symmetric}),
\begin{eqnarray}
S_{\sigma_i,\sigma_j}
&=&
\langle tr_{m+n}(\sigma_i X^{\dagger \otimes m}\otimes Y^{\dagger \otimes n})
tr_{m+n}(\sigma_j X^{\otimes m}\otimes Y^{\otimes n})\rangle
\nonumber \\
&=&\sum_{h\in H}tr_{m+n}(h\sigma_i h^{-1}\sigma_j)
\nonumber \\
&=&
\frac{N^{m+n}}{(m+n)!}\sum_{h\in H}
tr^{(r)}(\Omega_{m+n}h\sigma_i h^{-1}\sigma_j),
\end{eqnarray}
where $\sigma_i,\sigma_j\in S_{m+n}$ and   
$H=S_m\times S_n$. 
With the notation
(\ref{2pt_functions_new_notation}),    
one may write it as 
\begin{eqnarray}
S_{[\sigma_i]_H,[\sigma_j]_H}
&=&
N^{m+n}\frac{m!n!}{(m+n)!}
tr^{(r)}(\Omega_{m+n}[\sigma_i ]_H [\sigma_j]_H). 
\end{eqnarray}
Comparing to the three-point function (\ref{3pt_thetaTFT_symmetric}), 
we find that the $S_{\sigma_i,\sigma_j}$ can be interpreted to be 
the three-point function with the Omega factor put at one of the boundaries 
(see also \cite{1301.1980}). 
More explicitly, we have 
\begin{eqnarray}
S_{\sigma_i,\sigma_j}=N^{m+n}\frac{m!n!}{(m+n)!}
\sum_{k}\Xi_{ijk} 
N^{C_{\sigma_k}-(m+n)}.
\end{eqnarray}
Because $\Omega_{m+n}$ is a central element in 
the group algebra of $S_{m+n}$, 
the ordering of the three objects is not important.

\quad 

Next consider the way of labelling multi-traces in terms of Brauer elements.
Recall the 
two-point function (\ref{two-point_Brauer}):
\begin{eqnarray}
B_{b_i,b_j}&=&\langle tr_{m,n}(b_i X^{\otimes m}\otimes Y^{T \otimes n})
tr_{m,n}(b_j
X^{\dagger \otimes m}\otimes Y^{\ast \otimes n})\rangle
\nonumber \\
&=&
\sum_{h \in S_m\times S_n}
tr_{m,n}(hb_i h^{-1}b_j)
\nonumber \\
&=&
\sum_{\gamma}\sum_{h \in S_m\times S_n}t^{\gamma}\chi^{\gamma}(hb_i h^{-1}b_j).
\label{Brauer_correlator_character}
\end{eqnarray}
We can also write it as 
\begin{eqnarray}
B_{[b_i]_H,[b_j]_H}=\frac{1}{m!n!}
\sum_{\gamma}t^{\gamma}\chi^{\gamma}([b_i ]_H[b_j]_H).
\end{eqnarray}
The sum is over all Young diagrams $\gamma=(\gamma_+,\gamma_-)$ by means of 
(\ref{Brauer_Youngdiagram_sum}), 
where 
$\gamma_+$ and $\gamma_-$ are partitions of $m-k$ and $n-k$ respectively.
We now use the fact that 
the dimension of an irreducible representation $\gamma$ 
of 
the $GL(N)$ group can be expressed in terms of elements 
in $S_{m-k}\times S_{n-k}$ as
\begin{eqnarray}
t_{\gamma}=\frac{N^{m+n-2k}}{(m-k)!(n-k)!} 
\chi_{(\gamma_+,\gamma_-)}(\Omega_{m-k,n-k}). 
\end{eqnarray} 
The character of $S_{m-k}\times S_{n-k}$ is denoted by 
$\chi_{(\gamma_+,\gamma_-)}$, 
and $\Omega_{m-k,n-k}$ is a central element 
in the group algebra of $S_{m-k}\times S_{n-k}$ which is called 
coupled Omega factor \cite{GrossTaylor}. 
We do not use an explicit form of $\Omega_{m-k,n-k}$, but we should 
keep in mind that 
the $N$-dependence of $t_{\gamma}$is encoded in it, 
and it is not just the product $\Omega_{m-k}\times \Omega_{n-k}$. 
In order to combine $t^{\gamma}$ and $\chi^{\gamma}(hb_i h^{-1}b_j)$, 
we will 
express 
$\chi^{\gamma}(hb_i h^{-1}b_j)$ 
in terms of characters in $S_{m-k}\times S_{n-k}$ by
using  
the formulae (\ref{character_formula_brauer_1}) and 
(\ref{formula_new_Brauer_character}). 

Let us first consider the simplest case that both $b_i$ and $b_j$ 
are elements in the group algebra of $S_m\times S_n $ 
as an exercise before going to the general case.  
Suppose $b_i =b_i^+\otimes b_i^-$, $b_j =b_j^+\otimes b_j^-$, 
where $b_i^+$ and $b_j^+$ are elements in $S_m$ and 
$b_i^-$ and $b_j^-$ are elements in $S_n$.
For $h=\sigma\otimes \tau \in S_m\times S_n $,
\begin{eqnarray}
hb_i h^{-1}b_j
&=&(\sigma b_i^+ \sigma^{-1}b_j^+)\otimes 
(\tau b_i^-  \tau^{-1}b_j^- ) \in S_m\times S_n. 
\end{eqnarray}
Applying the formula (\ref{formula_new_Brauer_character}) 
to $\chi^{\gamma}(hb_i h^{-1}b_j)$, 
we obtain  
\begin{eqnarray}
\chi^{\gamma}(hb_i h^{-1}b_j)
&=&\frac{1}{(m-k)!(n-k)!}
\sum_{\sigma_2 \in S_{m-k},\tau_2\in S_{n-k}}
\Delta(\sigma b_i^+ \sigma^{-1}b_j^+,
\tau b_i^-  \tau^{-1}b_j^- ;\sigma_2,\tau_2)
\nonumber \\
&&
\times \chi_{(\gamma_+,\gamma_-)}(\sigma_2^{-1}\otimes \tau_2^{-1}), 
\end{eqnarray}
where 
\begin{eqnarray}
\chi_{(\gamma_+,\gamma_-)}(\sigma_2^{-1}\otimes \tau_2^{-1})
=\chi_{\gamma_+}(\sigma_2^{-1})
\chi_{\gamma_-}(\tau_2^{-1}).
\end{eqnarray}
Then the the two-point function 
can be written as 
\begin{eqnarray}
&&
B_{b_i,b_j}
\nonumber \\
&=&\sum_{\gamma}\sum_{h \in S_m\times S_n}t^{\gamma}\chi^{\gamma}(hb_i h^{-1}b_j)
\nonumber \\
&=&
\sum_{k=0}^{min(m,n)} \sum_{\gamma_+\vdash (m-k), \gamma_-\vdash (n-k)}
\sum_{\sigma\in S_m ,\tau \in S_n}
\frac{N^{m+n-2k}}{(m-k)!(n-k)!} 
\chi_{(\gamma_+,\gamma_-)}(\Omega_{m-k,n-k})
\nonumber \\
&&
\times 
\frac{1}{(m-k)!(n-k)!}
\sum_{\sigma_2 \in S_{m-k},\tau_2\in S_{n-k}}
\Delta(\sigma b_i^+ \sigma^{-1}b_j^+,
\tau b_i^-  \tau^{-1}b_j^- ;\sigma_2,\tau_2)
\chi_{(\gamma_+,\gamma_-)}(\sigma_2^{-1}\otimes \tau_2^{-1})
\nonumber \\
&=&
\sum_{k=0}^{min(m,n)}
\sum_{\sigma\in S_m ,\tau \in S_n}
N^{m+n-2k} \frac{1}{(m-k)!(n-k)!}
\nonumber \\
&&\times 
\sum_{\sigma_2 \in S_{m-k},\tau_2\in S_{n-k}}
\Delta(\sigma b_i^+ \sigma^{-1}b_j^+,
\tau b_i^-  \tau^{-1}b_j^- ;\sigma_2,\tau_2)
\delta_{m-k,n-k}(\Omega_{m-k,n-k}\sigma_2^{-1}\otimes \tau_2^{-1}), 
\label{two-point_brauer_rewriting_simplest}
\end{eqnarray}
where $\delta_{m-k,n-k}(\sigma\otimes \tau)$ is the delta function defined over 
the group algebra of $S_{m-k}\times S_{n-k}$. 

By the way in this case we know that 
the second line in 
(\ref{Brauer_correlator_character})
is factorised to give
\begin{eqnarray}
&&\sum_{h\in S_m\times S_n} tr_{m,n}(hb_i h^{-1}b_j)
\nonumber \\
&=&\sum_{\sigma\in S_m,\tau\in S_n} tr_{m}(\sigma b_i^+ \sigma^{-1}b_j^+)
tr_{n}(\sigma b_i^- \sigma^{-1}b_j^-)
\nonumber \\
&=&\sum_{\sigma\in S_m,\tau\in S_n}N^{m+n}
\delta_m(\Omega_m\sigma b_i^+ \sigma^{-1}b_j^+)
\delta_n(\Omega_n \sigma b_i^- \sigma^{-1}b_j^-). 
\label{factrisation_two-point}
\end{eqnarray}
Comparing (\ref{two-point_brauer_rewriting_simplest})
with (\ref{factrisation_two-point}) 
gives
an identity 
\begin{eqnarray}
&&
\delta_{m,n}(\Omega_m \sigma \otimes \Omega_n \tau)
=
\sum_k 
N^{-2k} \frac{1}{(m-k)!(n-k)!}
\nonumber \\
&& \qquad \times 
\sum_{\sigma_2 \in S_{m-k},\tau_2\in S_{n-k}}
\Delta(\sigma,\tau;\sigma_2,\tau_2)
\delta_{m-k,n-k}(\Omega_{m-k,n-k}\sigma_2^{-1}\otimes \tau_2^{-1}),
\end{eqnarray}
where $\sigma$ and $\tau$ are elements in $S_m$ and 
$S_n$ respectively.

Let us next study a general situation, 
where $b_i$ and $b_j$ are general elements in the Brauer algebra.  
We define 
$b_{ij}^{\alpha}:=\alpha b_i \alpha^{-1}b_j$. 
(We will use $\alpha$ for elements in $S_m\times S_n$ 
in stead of $h$ not to get an extra confusion with 
the number of contractions $h$.)
Using the formula 
(\ref{character_formula_brauer_1}),  
\begin{eqnarray}
\sum_{\alpha\in S_m\times S_n}\chi^{\gamma}(\alpha b_i \alpha^{-1}b_j)
=
\sum_{\alpha\in S_m\times S_n}
N^{z(b_{ij}^{\alpha})-h_{\alpha}}\chi^{\gamma}
(C^{\otimes h_{\alpha}}\otimes b_{ij}^{\alpha+}\otimes b_{ij}^{\alpha+}),
\end{eqnarray}
where 
$b_{ij}^{\alpha+}\otimes b_{ij}^{\alpha+}$ is an element in the group algebra of 
$S_{m-h_{\alpha}}\times S_{n-h_{\alpha}}$, and 
$z(b_{ij}^{\alpha})$ is the number of zero-cycles in $b_{ij}^{\alpha}$, and  
$h_{\alpha}$, $b_{ij}^{\alpha+}$, $b_{ij}^{\alpha-}$ are defined 
by this equation. 
We have added the subscript $\alpha$ to mean that they depend on $\alpha$. 

We will get an expression for $B_{b_i,b_j}$,
\begin{eqnarray}
B_{b_i,b_j}
&=&\sum_{\gamma}\sum_{\alpha \in S_m\times S_n}t^{\gamma}
\chi^{\gamma}(\alpha b_i \alpha^{-1}b_j)
\nonumber \\
&=&
\sum_{\gamma}\sum_{\alpha \in S_m\times S_n}t^{\gamma}
N^{z(b_{ij}^{\alpha})-h_{\alpha}}\chi^{\gamma}
(C^{\otimes h_{\alpha}}\otimes b_{ij}^{\alpha+}\otimes b_{ij}^{\alpha-})
\nonumber \\
&=&
\sum_{\alpha \in S_m \times  S_n}
\sum_{k=h_{\alpha}}^{min(m,n)} 
N^{m+n-2k} N^{z(b_{ij}^{\alpha})}\frac{1}{(m-k)!(n-k)!}
\nonumber \\
&& \times 
\sum_{\sigma_2 \in S_{m-k},\tau_2\in S_{n-k}}
\Delta(b_{ij}^{\alpha+},
b_{ij}^{\alpha-};\sigma_2,\tau_2)
\delta_{m-k,n-k}(\Omega_{m-k,n-k}\sigma_2^{-1}\otimes \tau_2^{-1}).
\end{eqnarray}
In this case we do not have a factorisation like 
(\ref{factrisation_two-point}). 
We can also write it as 
\begin{eqnarray}
B_{b_i,b_j}
&=&
\sum_{\alpha \in S_m \times  S_n}
\sum_{k=h_{\alpha}}^{min(m,n)} 
N^{m+n-2k} N^{z(b_{ij}^{\alpha})}
\nonumber \\
&&\times 
\frac{1}{((k-h_{\alpha})!(m-k)!(n-k)!)^2}
\sum_{\sigma_2 \in S_{m-k},\tau_2\in S_{n-k}}
\nonumber \\
&&\times 
\sum_{\sigma_1,\tau_1\in S_{k-h_{\alpha}}}
\eta^{S_{m-h_{\alpha}}}(b_{ij}^{\alpha+},\sigma_1\circ \sigma_2)
\eta^{S_{n-h_{\alpha}}}(b_{ij}^{\alpha-},\tau_1\circ \tau_2)
\eta^{S_{k-h_{\alpha}}}(\sigma_1^{-1}, \tau_1^{-1})
\nonumber \\
&&
\times 
\eta^{S_{m-k}\times S_{n-k}}(\Omega_{m-k,n-k},\sigma_2^{-1}\otimes \tau_2^{-1}).
\label{Brauer_twopt_decomposed}
\end{eqnarray}
\begin{figure}[t]
\begin{center}
\includegraphics[scale=0.6]{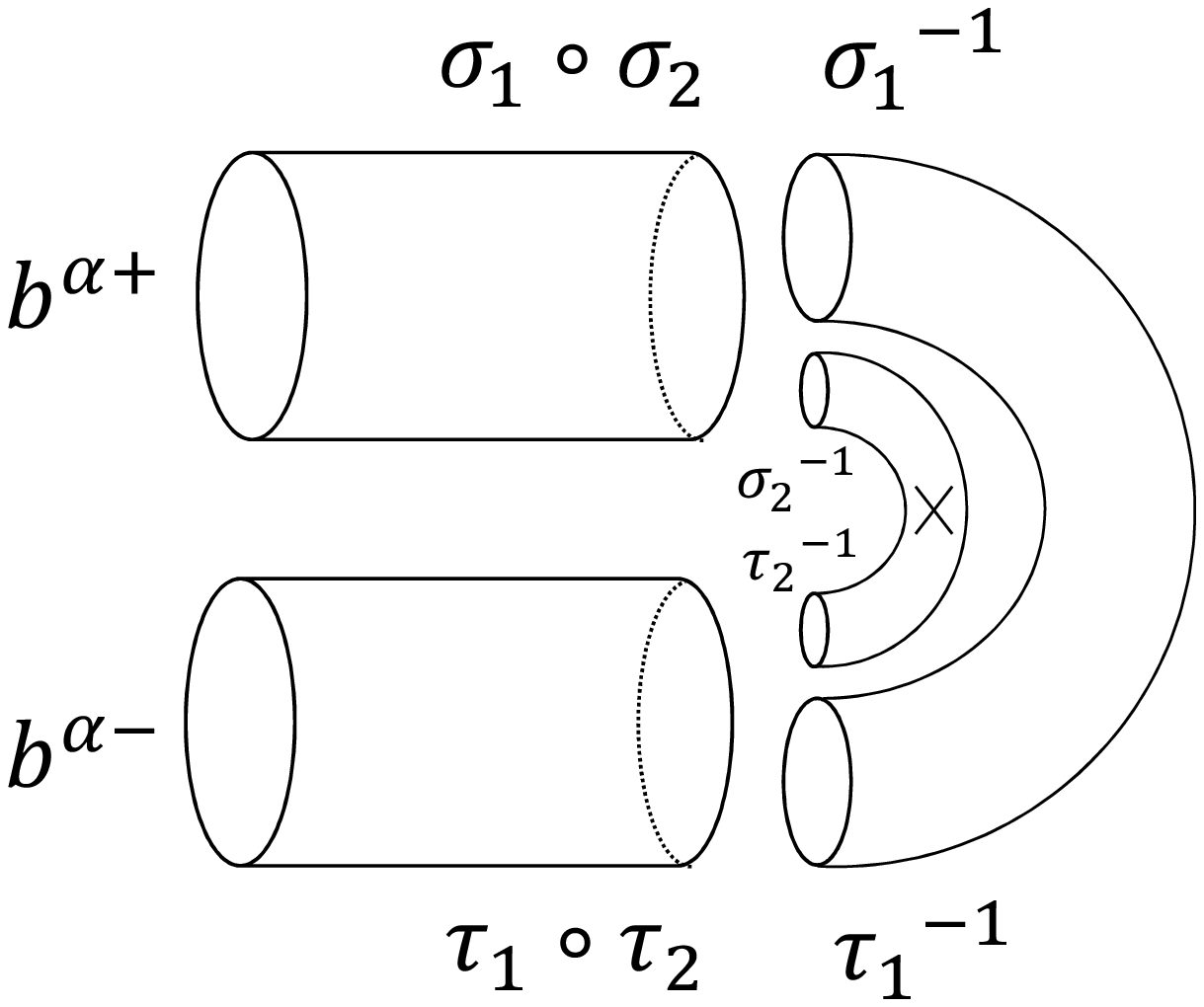}
\caption{
drawing of (\ref{Brauer_twopt_decomposed}): the {\it cross} 
represents the insertion of $\Omega_{m-k,n-k}$ 
} 
\label{fig:Brauertwopt_decomposed}
\end{center}
\end{figure}
A pictorial drawing is presented in figure \ref{fig:Brauertwopt_decomposed}.
Each piece in the above equation 
has an interpretation as 
a correlation function of the topological field theory based on the symmetric group. 
What we have done is almost the same as 
what we have done in subsection \ref{BrauerTFT_by_symmetricTFT}. 
Correlation functions 
expressed in terms of elements in  
the Brauer algebra have an interpretation as an assemblage of
correlation functions in terms of symmetric group elements.



\section{Some relations among the 2D quantum field theories}
\label{ref:relations_counting}
In this section we will explore some relations among the field theories 
we have shown, by exploiting a map $\Sigma$ between elements of the Brauer algebra 
and elements of the symmetric group,
\footnote{ 
Elements of the symmetric group $S_{m+n}$ are 
expressed diagrammatically 
by $m+n$ vertical edges between two horizontal lines where each horizontal line 
has $m+n$ points. 
Similarly 
elements of the walled Brauer algebra $B_N(m,n)$ 
are expressed 
by $m+n$ lines between the two horizontal lines 
with a vertical barrier separating the $m$ points from the $n$ points. 
Vertical edges do not cross the wall, and horizontal edges start and 
end on opposite sides of the wall. 
The map $\Sigma$ reflects the upper
right segment and the lower right segment into each other. 
For more details see section 3.3 in \cite{0709.2158}.
} 
which was exploited in \cite{0709.2158,0801.2061,0802.3662}.

Under the map,
elements of the Brauer algebra $B_N(m,n)$
are related to elements of the symmetric group $S_{m+n}$.
Suppose that a basis in $B_N(m,n)$ is related to a basis in $S_{m+n}$ 
by 
\begin{eqnarray}
\sigma_i=\Sigma (b_i), 
\end{eqnarray}
and the inverse map is denoted by 
\begin{eqnarray}
b_i=\Sigma^{-1}(\sigma_i). 
\end{eqnarray}
When two bases are related by the map in the above way, 
we can show 
\begin{eqnarray}
tr^{(r)}(b_ib^j)=tr^{(r)}(\sigma_i\sigma^j). 
\label{two_inner_product}
\end{eqnarray}
Note that $b^j$ is the dual basis obtained from a basis $b_i$ and 
$\sigma^j$ is the dual basis obtained from a basis $\sigma_i$, and 
the two dual bases are not related by the map, i.e. $\sigma^i\neq \Sigma (b^i)$. 
In fact we find that 
the both sides are equal to $\delta_i{}^j$, which come from the definition of 
the dual bases. 
Here we will give a proof of the equality (\ref{two_inner_product}) 
using the property of $\Sigma$, 
\begin{eqnarray}
tr^{(r)}(b_ib^j)
&=&tr_{m,n}(b_ib_j^{\ast})
\nonumber \\
&=&tr_{m+n}(\Sigma(b_i)\Sigma(b_j^{\ast}))
\nonumber \\
&=&tr_{m+n}(\Sigma(b_i)(\Sigma(b_j))^{-1}\Sigma(1^{\ast}))
\nonumber \\
&=&tr_{m+n}(\Sigma(b_i)(\Sigma(b_j))^{-1}\Omega_{m+n}^{-1})\frac{1}{N^{m+n}}
\nonumber \\
&=&\delta_{m+n}
(\Sigma(b_i)(\Sigma(b_j))^{-1})
\nonumber \\
&=&tr^{(r)}(\sigma_i\sigma^j),
\end{eqnarray}
where 
we have used the following equations found in \cite{0709.2158},
\begin{eqnarray}
&&
\Sigma(b^{\ast})=\Sigma(1^{\ast})(\Sigma(b))^{-1}
\nonumber \\
&&
\Sigma(1^{\ast})=\frac{1}{N^{m+n}}\Omega_{m+n}^{-1}.
\end{eqnarray}
Note that 
neither 
$tr^{(r)}(b_ib_j)=tr^{(r)}(\sigma_i\sigma_j)$ nor 
$tr^{(r)}(b^ib^j)=tr^{(r)}(\sigma^i\sigma^j)$ is satisfied. 
The relation (\ref{two_inner_product}) can be generalised to 
\begin{eqnarray}
tr^{(r)}(\Sigma^{-1}(\tau)b^j)=tr^{(r)}(\tau\sigma^j) 
\label{two_inner_product2}
\end{eqnarray}
for any element $\tau$ in the group algebra of $S_{m+n}$.

\quad 

We will consider the two-point function of the topological field theories 
obtained from the symmetric group $S_{m+n}$ and the Brauer algebra $B_N(m,n)$. 
The two correlation functions are related 
if 
boundary elements belong to the group algebra of 
$S_m\times S_n$, as we will show below. 
For $h_1=h_1^+\otimes h_1^-$,$h_2=h_2^+\otimes h_2^- \in S_m\times S_n$, 
using (\ref{two_inner_product2}),  
\begin{eqnarray}
\eta^{S_{m+n}}(h_1,h_2)
&=&\sum_{i}
tr^{(r)}(h_1\sigma_i h_2\sigma^i)
\nonumber \\
&=&\sum_{i}
tr^{(r)}(\Sigma^{-1}(h_1\sigma_i h_2) b^i)
\nonumber \\
&=&\sum_{i}
tr^{(r)}((h_1^+ \circ (h_2^-)^{-1}) \Sigma^{-1}(\sigma_i ) (h_2^+ \circ (h_1^-)^{-1})b^i)
\nonumber \\
&=&\sum_{i}
tr^{(r)}((h_1^+ \circ (h_2^-)^{-1}) b_i (h_2^+ \circ (h_1^-)^{-1})b^i)
\nonumber \\
&=&
\eta^{B_N(m,n)}(h_1^+ \circ (h_2^-)^{-1},h_2^+ \circ (h_1^-)^{-1}), 
\label{relation_two_cylinders}
\end{eqnarray}
where 
we have used the following equation
\begin{eqnarray}
\Sigma^{-1}(h_1\sigma h_2)=
(h_{1}^+\circ (h_{2}^-)^{-1})\Sigma^{-1}(\sigma)(h_{2}^+\circ (h_{1}^-)^{-1}), 
\label{an_equation_of_Sigma}
\end{eqnarray}
which come from the definition of $\Sigma$. 
Note that we have used the same symbol for the regular representation of 
$S_{m+n}$ and that of $B_N(m,n)$. 

Gluing two boundaries of the cylinders, we have
\begin{eqnarray}
\sum_{h\in S_m\times S_n} \eta^{S_{m+n}}(h,h^{-1})
&=&
\sum_{h^+\in S_m,h^-\in S_n}
\eta^{B_N(m,n)}(h^+ \circ h^-,(h^+)^{-1} \circ (h^-)^{-1})
\nonumber \\
&=&
\sum_{h\in S_m\times S_n}
\eta^{B_N(m,n)}(h,h^{-1}). 
\label{relation_two_cylinders_glued}
\end{eqnarray}
One find that 
gluing two ends of 
the cylinder with boundary elements restricted to $S_m\times S_n$
gives 
the torus given in section \ref{sec:new_theory_subgroup},
\begin{eqnarray}
&&
\sum_{h\in S_m\times S_n} \eta^{S_{m+n}}(h,h^{-1})
=\theta^{S_{m+n}}{}_i{^i}
\nonumber \\
&&
\sum_{h\in S_m\times S_n} \eta^{B_N(m,n)}(h,h^{-1})
=\theta^{B_N(m,n)}{}_i{^i}. 
\end{eqnarray}
Therefore 
(\ref{relation_two_cylinders_glued}) means 
\begin{eqnarray}
\theta^{S_{m+n}}{}_i{^i}
=\theta^{B_N(m,n)}{}_i{^i}. 
\label{two_torus_equivalence}
\end{eqnarray}
Taking into account that the partition function of a torus 
gives the dimension of the vector space, 
this equation states that 
the vector space of 
$[\sigma_i]_H$ and the vector space of  
$[b_i]_H$ have the same dimension.

In fact the result itself has been known 
from 
the counting of gauge invariant operators at large $N$. 
Let the number of gauge invariant operators built from 
$m$ copies of $X$ and $n$ copies of $Y$ be $N(m,n)$. 
There are several ways to obtain this. 
If we use 
the restricted Schur basis
(\ref{restricted_schur_two-point}), we obtain 
$N(m,n)=\sum_{R,A}(M^R_A)^2$ \cite{0801.2061,0810.4217},
which is equal to $\theta^{S_{m+n}}{}_i{^i}$.  
If we use 
the Brauer basis 
(\ref{two-point_Brauer}),
we have $N(m,n)=\sum_{\gamma,A}(M^{\gamma}_A)^2$ 
\cite{0709.2158,0911.4408}, 
which is equal to $\theta^{B_N(m,n)}{}_i{^i}$. 
Thus (\ref{two_torus_equivalence}) indeed is an expected equation 
from the counting of multi-trace operators at large $N$. 
It is interesting that 
\begin{eqnarray}
\sum_{R,A}(M^R_A)^2=\sum_{\gamma,A}(M^{\gamma}_A)^2
\label{matching_counting}
\end{eqnarray}
has been derived 
as a consequence of the relation (\ref{relation_two_cylinders}) between two 
topological field theories.


\section{Discussions}
\label{discussions}

Having studied two-dimensional (almost topological) field theories 
related to the multi-matrix models, it is now good to 
get back to the equations in (\ref{2pt_functions_new_notation}). 
Let us first see the left-hand side. It is 
given in the $U(N)$ gauge theory language, and 
gauge invariant quantities are integrated.  
Multi-trace operators are characterised by the invariance under the gauge transformation,
\begin{eqnarray}
X\rightarrow gXg^{-1}, \quad Y\rightarrow gYg^{-1}.
\end{eqnarray}
(Enhanced symmetries at the free theory 
of ${\cal N}=4$ SYM
were discussed in \cite{0807.3696}.)
On the other hand, 
the equations in 
the right-hand side are completely expressed 
in terms of the symmetric group or the walled Brauer algebra. 
The quantities inside the trace,  
$[\sigma]$, $[\sigma]_H$ and $[b]_H$,  are characterised by the 
invariance under the following gauge transformation,
\begin{eqnarray}
\sigma \rightarrow h\sigma h^{-1}, \quad b\rightarrow hbh^{-1}
\quad (h\in S_m \times S_n) .
\end{eqnarray}
One might regard the equations in (\ref{2pt_functions_new_notation}) 
as an analogue of the GKPW relation \cite{9802109,9802150}. 
Mathematical manipulations behind the equality can be understood 
from the Schur-Weyl duality. 
Frobenius algebras are algebras of these gauge invariant quantities, 
and the bilinear forms (the Frobenius forms) 
are projection operators onto gauge invariant quantities. 
These Frobenius algebras are noncommutative 
except for the case of the one-complex matrix model. 
The noncommutativity is related to the existence of 
the multiplicity indices $\mu,\nu$ on the restricted characters, 
which are originally introduced to describe open strings on giant gravitons 
\cite{0411205,0701066}. 
Noncommutative Frobenius algebras would play a role 
in the description of two-dimensional field theories 
with a certain structure relevant for the noncommutativity.   
It would be interesting to use axiomatic 
notions of noncommutative Frobenius algebras
to understand 
what kind of field theories are described reflecting the noncommutative nature.
For example see \cite{0010269,math/0202164,0609042,math0508349} for references of 
the direction.
Even apart from the connection to ${\cal N}=4$ SYM, 
noncommutative Frobenius algebras themselves seem to be
an interesting subject to learn.

In subsection \ref{BrauerTFT_by_symmetricTFT} and section \ref{N=4SYM_2pt_TFT},  
we have decomposed correlation functions of 
two-dimensional quantum field theories obtained from 
walled Brauer algebras into correlation functions of 
two-dimensional quantum field theories obtained from 
symmetric groups, with exploiting the property that 
the character of the walled Brauer algebra can be expressed in terms of 
the character of the symmetric group $S_{m-k}\times S_{n-k}$. 
From the link between permutations and coverings,  
walled Brauer algebras can also be interesting mathematical tools 
incorporate two kinds of maps, i.e.  
a holomorphic map from $m$ worldsheets 
to the target and an anti-holomorphic map from $n$ worldsheets. 
The complete large $N$ expansion of two-dimensional Yang-Mills 
is a well-known example \cite{GrossTaylor,9402107,9411210}. 
A non-holomorphic extension 
of 
\cite{1002.1634,1104.2386}
exploiting walled Brauer algebras 
might be a possible future direction. 
 
Throughout this paper, we have assumed that 
$N$ is large compared to the number of fields involved in 
multi-traces (i.e. $n<N$ for the one-matrix model, and $m+n<N$ 
for the two-matrix model - we call this large $N$). 
If the number of fields exceed the bound $N$ (i.e. $n>N$ or $m+n>N$ - we call this small $N$), 
it is good to use the diagonal operator basis, 
$\langle O_R O_S \rangle =N^n tr^{(r)}(\Omega_n p_Rp_S)$
instead of (\ref{2pt_functions_new_notation}). 
In the AdS/CFT correspondence, gauge invariant operators are considered to 
be dual to D-branes (called giant gravitons) or geometries if 
the number of fields are comparable to $N$ or $N^2$, 
and considering representation bases clarifies the correspondence between 
gauge theory operators and string states.  
It would be interesting to 
study a two-dimensional interpretation 
of representation bases, with focusing the role of 
Frobenius algebras.

\vspace{0.4cm}

\noindent
{\bf Acknowledgements}
I would like to thank Sanjaye Ramgoolam for valuable discussions. 
E-mail conversations with him motivated me to initiate this work.

\vspace{0.4cm}


\appendix 
\renewcommand{\theequation}
{\Alph{section}.\arabic{equation}}

\section{Symmetric groups, Brauer algebras and Schur-Weyl duality}
\label{SWdual_symmetric_Brauer}
\setcounter{equation}{0}

In this section we will make a brief 
introduction of 
symmetric groups and walled Brauer algebras, focusing on the role of 
Schur-Weyl duality in the description of multi-traces. 

Let $V$ be an $N$-dimensional vector space over $\mathbb{C}$ 
on which an $N\times N$ matrix $X$ is supposed to act. 
The tensor product $X^{\otimes n}=X\otimes \cdots \otimes X$ can be viewed 
as an operator acting on the tensor space
$V^{\otimes n}$. 
We define an action of the symmetric group $S_{n}$
as permuting $n$ vector spaces,
\begin{eqnarray}
\sigma (v_1\otimes \cdots \otimes v_n)=
v_{\sigma^{-1}(1)}\otimes \cdots \otimes v_{\sigma^{-1}(n)}.
\end{eqnarray}
In addition to this 
the $GL(N)$ group acts 
in the standard way 
on it by the simultaneous matrix multiplication,
\begin{eqnarray}
g (v_1\otimes \cdots \otimes v_n)=
gv_1\otimes \cdots \otimes gv_n .
\end{eqnarray}
These two actions can be shown to commute, 
and the Schur-Weyl duality says that the tensor space is 
decomposed into the direct sum of irreducible representations for these groups as 
\begin{eqnarray}
V^{\otimes n}=\bigoplus_{R\vdash n}
V_{R}^{GL(N)}\otimes V_{R}^{S_{n}}.  
\label{SW_duality}
\end{eqnarray}
The sum is taken for 
all Young diagrams with 
$n$ boxes satisfying $c_{1}(R)\le N$, 
where $c_{1}(R)$ is the number of rows in $R$. 
The projection operator $p_{R}$ 
of an irreducible representation $R$ 
can be introduced as  
an element 
in the group algebra of $S_{n}$,
\begin{eqnarray}
p_{R}=\frac{d_{R}}{n!}\sum_{\sigma\in S_{n}}\chi_{R}(\sigma^{-1})\sigma, 
\label{projector_S_n_single}
\end{eqnarray}
which acts on the tensor space as
\begin{eqnarray}
p_{R}V^{\otimes n}=
V_{R}^{GL(N)}\otimes V_{R}^{S_{n}}.  
\end{eqnarray}
We have denoted 
by $d_R$
the dimension of an irreducible representation $R$ of the symmetric group, 
and let  
$t_R$ be 
the dimension of an irreducible representation $R$ of the $GL(N)$ group. 

Let $tr_n$ be a trace 
over the tensor space $V^{\otimes n}$. 
From the Schur-Weyl duality (\ref{SW_duality}), 
we have 
\begin{eqnarray}
tr_n(\sigma)
=\sum_R t_R \chi_R(\sigma)
\label{swdual_trace}
\end{eqnarray}
for an element $\sigma$ in the symmetric group. 
If we introduce 
$C_{\sigma}$, the number of cycles in the permutation $\sigma$, 
we have $tr_n(\sigma)=N^{C_{\sigma}}$.
The orthogonality relation of the characters leads to 
\begin{eqnarray}
t_R 
=\frac{N^n}{n!}\chi_R(\Omega_n),
\label{dimR_symmetric_data}
\end{eqnarray}
where 
$\Omega_n$ is 
a central element in the group algebra of $S_n$ called Omega factor
 \begin{eqnarray}
\Omega_n=
 \sum_{\sigma\in S_n}\sigma N^{C_{\sigma}-n}.
\end{eqnarray}
By substituting (\ref{dimR_symmetric_data}) back to (\ref{swdual_trace}), 
the trace of 
an element $\tau$ 
in the group algebra of $S_n$
can be written as 
\begin{eqnarray}
tr_n(\tau)&=&\frac{N^n}{n!}
\sum_{R\vdash n} \chi_R(\Omega_n) \chi_R(\tau)
\nonumber \\
&=&\frac{N^n}{n!}
\sum_{R\vdash n} d_R \chi_R(\Omega_n\tau)
\nonumber \\
&=&N^n
\delta_n(\Omega_n\tau),
\label{tr_n_tau_evaluate}
\end{eqnarray}
where we have introduced the delta function defined  
over the group algebra of $S_n$
by $\delta_n(\sigma)=1$ if $\sigma=1$ and 
$0$ otherwise,
\begin{eqnarray}
\delta_n(\sigma)=\frac{1}{n!}\sum_{R\vdash n} d_R\chi_R(\sigma). 
\label{def_delta_symmetric_group}
\end{eqnarray}
In order to derive (\ref{tr_n_tau_evaluate}), we have assumed $N>n$. 
When this is not satisfied, only irreducible representations satisfying 
the constraint $c_1(R)\le N$ are summed in the first line 
and the second line 
of (\ref{tr_n_tau_evaluate}). 
On the other hand 
all 
Young diagrams with $n$ boxes are summed in (\ref{def_delta_symmetric_group}).

\quad 

We next consider 
the vector space that are relevant for the description of 
multi-trace operators made from 
two matrices $X$ and $Y$. 
When  
$m$ copies of $X$ and $n$ copies of $Y$ are considered, the tensor space we will consider is 
$V^{\otimes (m+n)}$, and 
the Schur-Weyl duality claims
\begin{eqnarray}
V^{\otimes (m+n)}=\bigoplus_{R\vdash (m+n)}
V_{R}^{GL(N)}\otimes V_{R}^{S_{m+n}}.  
\label{SW_duality2}
\end{eqnarray}
We now consider 
the decomposition of the irreducible representation $R$ 
of $S_{m+n}$ into irreducible representations of 
the subgroup $S_{m}\times S_n$, 
\begin{eqnarray}
V_{R}^{S_{m+n}}=
\bigoplus_{A}M^R_A \hspace{0.1cm} V_{A}^{S_{m}\times S_{n}}.
\label{Sm+n_SmSn}
\end{eqnarray}
On restricting to the subgroup, some copies of $A$ appear. 
The number of times $A$ appears in $R$ is denoted by $M^R_A$, 
which is given by 
the Littlewood-Richardson coefficient 
\begin{eqnarray}
M^R_A=g(\alpha,\beta;R) \quad A=(\alpha,\beta).
\label{multiplicity_symmetric}
\end{eqnarray} 
We now define an operator 
$P^R_{A,\mu\nu}$ playing a role
under the decomposition. 
Here $\mu,\nu$ run over $1,\cdots ,M^R_A$, labelling which copy of $A$ we are using. 
If the multiplicity is trivial, $P^R_{A}$ is the projection operator onto 
the irreducible representation $A$ inside the $R$. 
If the multiplicity is non-trivial, 
$P^R_{A,\mu\nu}$ is an intertwiner mapping 
the $\nu$-th copy of the representations $A$ to 
the $\mu$-th copy of the representations $A$. 
It satisfies 
\begin{eqnarray}
P^{R}_{A,\mu\nu}
P^{R^{\prime}}_{A^{\prime},\mu^{\prime}\nu^{\prime}}
=\delta^{R R^{\prime}}
\delta_{A A^{\prime}}
\delta_{\nu \mu^{\prime}}
P^{R}_{A,\mu \nu^{\prime}}. 
\label{product_P_RAij}
\end{eqnarray}
In terms of elements in the symmetric group $S_{m+n}$, 
the operator $P^R_{A,\mu\nu}$
can be explicitly written as 
\begin{eqnarray}
P^{R}_{A,\mu\nu}
=\frac{d_R}{(m+n)!}\sum_{\sigma\in S_{m+n}}\chi^R_{A,\nu\mu}(\sigma)\sigma^{-1},
\label{intertwiner_symmetric_group}
\end{eqnarray}
where $\chi^R_{A,\nu\mu}(\sigma)$ is a quantity 
called restricted character \cite{0411205,0701066,0801.2061}. 
This can be computed by the trace of the matrix $\sigma$
in the representation $R$, but the trace is only over 
the subspace $A$ appearing in 
the $(\mu,\nu)$ component of the $M^R_A \times M^R_A$ matrix.
The sum of the diagonal copies of all possible irreducible representations 
in $S_m\times S_n$ inside the representation $R$ 
gives rise to the usual character 
$\chi^R(\sigma)=\sum_{A,\mu}\chi^R_{A,\mu\mu}(\sigma)$, and  
the projector of 
an irreducible representation $R$ is given by 
\begin{eqnarray}
P^{R}=\sum_{A,\mu}P^{R}_{A,\mu\mu}. 
\end{eqnarray}
The intertwiner operator has the symmetry, 
\begin{eqnarray}
hP^{R}_{A,\mu\nu}=P^{R}_{A,\mu\nu}h \quad (h \in S_m\times S_n).
\end{eqnarray}
But 
elements in $S_{m+n}/S_m\times S_n$ do not commute with it. 
Due to this, the cyclicity of the restricted character works for 
elements in the subalgebra,
\begin{eqnarray}
\chi^R_{A,\mu\nu}(h\sigma )=\chi^R_{A,\mu\nu}(\sigma h)
\quad (h \in S_m\times S_n).
\label{restricted_symmetry}
\end{eqnarray}

\vspace{0.4cm}

Let us next consider 
the mixed tensor space $V^{\otimes m}\otimes \bar{V}^{\otimes n}$ by 
including the complex conjugate space $\bar{V}$. 
This is relevant for the description of multi-trace operators 
made out of $X$ and $X^{\dagger}$, or $X$ and $Y^{T}$.  
Similar to the previous cases, 
we can consider two commuting actions on this space, resulting in the following 
Schur-Weyl duality, 
\begin{eqnarray}
V^{\otimes m}\otimes \bar{V}^{\otimes n}
=\bigoplus_{\gamma}
V_{\gamma}^{GL(N)}\otimes V_{\gamma}^{B_{N}(m,n)}. 
\label{SWdual_Brauer}
\end{eqnarray}
Here $B_{N}(m,n)$ is the walled Brauer algebra \cite{Stembridge,Koike,Turaev,BCHLLS}. 
(This algebra is sensitive to $N$, which is a big difference from 
the group algebra of the symmetric group.) 
The $\gamma$ is a set of two Young diagrams ($\gamma_+,\gamma_-$), 
where $\gamma_+$ is a Young diagram with $m-k$ boxes and 
$\gamma_-$ is a Young diagram with $n-k$ boxes.
The $k$ is an integer in $0\le k \le min(m,n)$. 
The sum over $\gamma$ in (\ref{SWdual_Brauer}) is constrained by
$c_1(\gamma_+)+c_1(\gamma_-)\le N$. 

From the Schur-Weyl duality, 
we have 
\begin{eqnarray}
tr_{m,n}(b)=\sum_{\gamma}t_{\gamma} \chi^{\gamma}(b) \quad 
(b\in B_N(m,n))
\label{tracemn_Brauer}
\end{eqnarray}
where $t_{\gamma}$ is the dimension of an irreducible representation $\gamma$ 
of the $GL(N)$ group, 
and $\chi^{\gamma}(b)$ is the character of an irreducible representation $\gamma$
of the Brauer algebra. The $tr_{m,n}$ denotes a trace over the mixed tensor space. 
Let $d_{\gamma}$ be  
the dimension of $\gamma$ 
in the Brauer algebra. 
We have a formula of $t_{\gamma}$ 
using elements in $S_m\times S_n$,
\begin{eqnarray}
t_{\gamma} =\frac{N^{m+n-2k}}{(m-k)!(n-k)!}
\chi_{(\gamma_+,\gamma_-)}(\Omega_{m-k,n-k}),
\end{eqnarray}
where $\chi_{(\gamma_+,\gamma_-)}(\sigma\otimes \tau)$
is the character of $S_{m-k}\times S_{n-k}$, 
and $\Omega_{m-k,n-k}$ is a central element in the group algebra of 
$S_{m-k}\times S_{n-k}$ called 
coupled Omega factor \cite{GrossTaylor}. 
A formula to express $\Omega_{m,n}^{-1}$ in terms of $\Omega_{m+n}^{-1}$ 
is given in \cite{0802.3662}. 
We have the following formula of $d^{\gamma}$, 
\begin{eqnarray}
d^{\gamma}=\frac{m!n!}{(m-k)!(n-k)!k!}d_{\gamma_+}d_{\gamma_-},
\label{Brauer_dimension_symmetric}
\end{eqnarray}
where $d_{\gamma_+}$ and $d_{\gamma_-}$ are
the dimensions of $S_{m-k}$ and $S_{n-k}$ respectively.

We can construct the projection operator $P^{\gamma}$ as\footnote{
The expression is valid for $m+n\le N$.}
\begin{eqnarray}
P^{\gamma}
=d^{\gamma}\sum_i \chi^{\gamma}(b^{i})b_i,
\label{projector_central_Brauer}
\end{eqnarray}
where $b^i$ is the dual basis in 
(\ref{Brauer_dual_basis}). 
The Schur-Weyl duality asserts
\begin{eqnarray}
P^{\gamma}V^{\otimes m}\otimes \bar{V}^{\otimes n}
=
V_{\gamma}^{GL(N)}\otimes V_{\gamma}^{B_{N}(m,n)}. 
\end{eqnarray}

Because the walled Brauer algebra $B_{N}(m,n)$ contains 
the group algebra of $S_{m}\times S_{n}$, 
which we denote by $\mathbb C(S_{m}\times S_{n})$, 
we have a decomposition similar to (\ref{Sm+n_SmSn}) 
\begin{eqnarray}
V_{\gamma}^{B_{N}(m,n)}=
\bigoplus_{A}M^{\gamma}_A \hspace{0.1cm} V_{A}^{\mathbb C(S_{m}\times S_{n})}, 
\end{eqnarray}
Here the multiplicity associated with the decomposition
is given by 
\begin{eqnarray}
M^{\gamma}_A=\sum_{\delta\vdash k}g(\gamma_+,\delta;\alpha)
g(\gamma_-,\delta;\beta), \quad A=(\alpha,\beta).
\label{multiplicity_Brauer}
\end{eqnarray} 
We can introduce an operator $Q^{\gamma}_{A,\mu\nu}$ as an element 
in the Brauer algebra \cite{0709.2158,0807.3696} that satisfies 
\begin{eqnarray}
Q^{\gamma}_{A,\mu\nu}
Q^{\gamma^{\prime}}_{A^{\prime},\mu^{\prime}\nu^{\prime}}
=\delta^{\gamma \gamma^{\prime}}
\delta_{A A^{\prime}}
\delta_{\nu \mu^{\prime}}
Q^{\gamma}_{A,\mu \nu^{\prime}}. 
\label{producQ}
\end{eqnarray}
The role of the operator $Q^{\gamma}_{A,\mu\nu}$ is completely the same as  
$P^{R}_{A,\mu\nu}$. 
Introducing 
the restricted character of the walled Brauer algebra, we have  
\begin{eqnarray}
Q^{\gamma}_{A,\mu\nu}
=d^{\gamma}\sum_{i}\chi^{\gamma}_{A,\nu\mu}(b^i)b_i .
\label{explicit_form_Q-operator}
\end{eqnarray} 
The relation between $P^{\gamma}$ and $Q^{\gamma}_{A,\mu\nu}$ is 
\begin{eqnarray}
P^{\gamma}=\sum_{A,\mu}Q^{\gamma}_{A,\mu\mu}. 
\end{eqnarray}
The intertwiner and the restricted character have the symmetry 
\begin{eqnarray}
&&
hQ^{\gamma}_{A,\mu\nu}h^{-1}=Q^{\gamma}_{A,\mu\nu},\nonumber \\
&&
\chi^{\gamma}_{A,\mu\nu}(hb h^{-1})=\chi^{\gamma}_{A,\mu\nu}(b)
\quad (h \in S_m\times S_n).
\label{restricted_Brauer_symmetry}
\end{eqnarray}

There is 
another way of using the Schur-Weyl duality in the description of 
multi-traces \cite{0711.0176,0910.2170}. 

\section{Orthogonality relations}
\setcounter{equation}{0}
\label{sec:formula_orthogonality}

In this section we summarise orthogonality relations of 
representations and 
characters of the symmetric group and the walled Brauer algebra.

Orthogonality relations of the symmetric group $S_n$ are 
\begin{eqnarray}
\frac{1}{n!}
\sum_{\sigma\in S_{n}}D^{R}(\sigma)_{ij}D^{S}(\sigma^{-1})_{kl}
=\frac{1}{d_R}\delta_{il}\delta_{jk}\delta_{RS},
\end{eqnarray}
\begin{eqnarray}
\frac{1}{n!}\sum_{\sigma\in S_n}\chi_{R}(\sigma)\chi_{S}(\sigma^{-1})
=\delta_{RS},
\label{character_sum_sigma}
\end{eqnarray}
\begin{eqnarray}
\frac{1}{n!}\sum_{\sigma\in S_n}\chi_{R}(\sigma)\chi_{S}(\sigma^{-1}\tau)
=\frac{1}{d_R}\chi_R(\tau)
\delta_{RS},
\label{character_sum_sigma_more}
\end{eqnarray}

\begin{eqnarray}
\frac{1}{d_R}\chi_{R}(\sigma)\chi_{R}(\tau)
=
\frac{1}{n!}
\sum_{\rho\in S_n}\chi_R(\rho\sigma \rho^{-1}\tau),
\label{character_combine_useful}
\end{eqnarray}
\begin{eqnarray}
\sum_{R\vdash n}\chi_{R}(\sigma)\chi_{R}(\tau)
=\sum_{\rho\in S_n}\delta_{n}(\rho\sigma \rho^{-1}\tau).
\label{character_sum_R_formula}
\end{eqnarray}

We have similar formulae for 
the restricted characters.  
For the symmetric group $S_{m+n}$ with the restriction to $S_m\times S_n$ considered, 
we have  
\begin{eqnarray}
\frac{1}{(m+n)!}
\sum_{\sigma\in S_{m+n}}\chi^{R}_{A,\mu\nu}(\sigma)
\chi^{R^{\prime}}_{A^{\prime},\mu^{\prime}\nu^{\prime}}(\sigma^{-1})
=\frac{d_A}{d_R}
\delta_{RR^{\prime}}
\delta_{AA^{\prime}}
\delta_{\mu\nu^{\prime}}
\delta_{\nu\mu^{\prime}},
\label{orthogonality_restricted_character_symmetric}
\end{eqnarray}
This formula is consistent with 
(\ref{character_sum_sigma}) due to $d_R=\sum_A d_A M^{R}_A$. We also have 
\begin{eqnarray}
\frac{1}{(m+n)!}
\sum_{\sigma\in S_{m+n}}\chi^{R}_{A,\mu\nu}(\sigma)
\chi^{R^{\prime}}_{A^{\prime},\mu^{\prime}\nu^{\prime}}(\sigma^{-1}\tau)
=\frac{1}{d_R}\chi^R_{A\mu\nu^{\prime}}(\tau)
\delta_{RR^{\prime}}
\delta_{AA^{\prime}}
\delta_{\mu^{\prime}\nu},
\label{orthogonality_restricted_character_symmetric_more1}
\end{eqnarray}
\begin{eqnarray}
\frac{1}{(m+n)!}
\sum_{\sigma\in S_{m+n}}\chi^{R}_{A,\mu\nu}(\sigma)
\chi^{R^{\prime}}_{A^{\prime},\mu^{\prime}\nu^{\prime}}(\tau\sigma^{-1})
=\frac{1}{d_R}\chi^R_{A\mu^{\prime}\nu}(\tau)
\delta_{RR^{\prime}}
\delta_{AA^{\prime}}
\delta_{\mu\nu^{\prime}},
\label{orthogonality_restricted_character_symmetric_more2}
\end{eqnarray}
\begin{eqnarray} 
\sum_{A,\mu,\nu}
\frac{1}{d_A}\chi^{R}_{A,\mu\nu}(\sigma_1)
\chi^{R}_{A,\nu\mu}(\sigma_2)
=
\frac{1}{m!n!}
\sum_{h\in S_m\times S_n}
\chi^{R}(h\sigma_1 h^{-1} \sigma_2).
\label{Qoperator_characer_combine_symmetric} 
\end{eqnarray} 
If we introduce the dual basis $\sigma^i=\frac{1}{n!}\sigma_i^{-1}$, 
factorials disappear from these formulae. 

For the walled Brauer algebra $B_N(m,n)$, we have 
\begin{eqnarray}
\sum_{b \in B_N(m,n)} 
D^{\gamma}(b)_{ij}D^{\gamma^{\prime}}(b^{\ast})_{kl}
=\frac{1}{t^{\gamma}} \delta_{il}\delta_{jk} \delta^{\gamma \gamma^{\prime}},
\label{orthogonality_rep_brauer}
\end{eqnarray}
\begin{eqnarray}
\sum_{b \in B_N(m,n)} 
\chi^{\gamma}(b)\chi^{\gamma^{\prime}}(b^{\ast})
=\frac{d^{\gamma}}{t^{\gamma}} \delta^{\gamma \gamma^{\prime}},
\label{orthogonality_character_brauer}
\end{eqnarray}
\begin{eqnarray}
\sum_{b \in B_N(m,n)} 
\chi^{\gamma}(b)\chi^{\gamma^{\prime}}(b^{\ast}c)
=\frac{1}{t^{\gamma}} \delta^{\gamma \gamma^{\prime}}
\chi^{\gamma}(c),
\label{orthogonality_character_brauer_more}
\end{eqnarray}
\begin{eqnarray}
\sum_{\gamma}
\chi^{\gamma}(b)\chi^{\gamma}(c)
=\sum_{d \in B_N(m,n)} 
t^{\gamma}
\chi^{\gamma}(dbd^{\ast}c),
\label{twocharacters_combine}
\end{eqnarray}

\begin{eqnarray}
\sum_{b \in B_N(m,n)} 
\chi^{\gamma}_{A,\mu\nu}(b)
\chi^{\gamma^{\prime}}_{A^{\prime},\mu^{\prime}\nu^{\prime}}(b^{\ast})
=\frac{d_A}{t^{\gamma}}
\delta_{\gamma\gamma^{\prime}}
\delta_{AA^{\prime}}
\delta_{\mu\nu^{\prime}}
\delta_{\mu^{\prime}\nu},
\label{orthogonality_restricted_character}
\end{eqnarray}

\begin{eqnarray}
\sum_{b \in B_N(m,n)} 
\chi^{\gamma}_{A,\mu\nu}(b)
\chi^{\gamma^{\prime}}_{A^{\prime},\mu^{\prime}\nu^{\prime}}(b^{\ast} c)
=\frac{1}{t^{\gamma}}
\delta_{\gamma\gamma^{\prime}}
\delta_{AA^{\prime}}
\chi^{\gamma}_{A,\mu\nu^{\prime}}(c)
\delta_{\mu^{\prime}\nu},
\label{orthogonality_restricted_character_Brauer_more1}
\end{eqnarray}
\begin{eqnarray}
\sum_{b \in B_N(m,n)} 
\chi^{\gamma}_{A,\mu\nu}(b)
\chi^{\gamma^{\prime}}_{A^{\prime},\mu^{\prime}\nu^{\prime}}(cb^{\ast} )
=\frac{1}{t^{\gamma}}
\delta_{\gamma\gamma^{\prime}}
\delta_{AA^{\prime}}
\chi^{\gamma}_{A,\mu^{\prime}\nu}(c)
\delta_{\mu\nu^{\prime}},
\label{orthogonality_restricted_character_Brauer_more2}
\end{eqnarray}
\begin{eqnarray} 
\sum_{A,\mu,\nu}
\frac{1}{d_A}\chi^{\gamma}_{A,\mu\nu}(b_1)
\chi^{\gamma}_{A,\nu\mu}(b_2)
=
\frac{1}{m!n!}\sum_{h\in S_m\times S_n}
\chi^{\gamma}(hb_1 h^{-1} b_2).
\label{Qoperator_characer_combine} 
\end{eqnarray} 

The orthogonality relations in terms of 
the dual basis $b^i=g^{ij}b_j$ are as follows:
\begin{eqnarray}
\sum_{i} 
D^{\gamma}(b_i)_{ij}D^{\gamma^{\prime}}(b^{i})_{kl}
=\frac{1}{d^{\gamma}} \delta_{il}\delta_{jk} \delta^{\gamma \gamma^{\prime}},
\label{orthogonality_rep_brauer_dual}
\end{eqnarray}
\begin{eqnarray}
\sum_{i} 
\chi^{\gamma}(b_i)\chi^{\gamma^{\prime}}(b^{i})
= \delta^{\gamma \gamma^{\prime}},
\label{orthogonality_character_brauer_dual}
\end{eqnarray}
\begin{eqnarray}
\sum_{i} 
\chi^{\gamma}(b_i)\chi^{\gamma^{\prime}}(b^{i}c)
=\frac{1}{d^{\gamma}} \delta^{\gamma \gamma^{\prime}}
\chi^{\gamma}(c),
\label{orthogonality_character_brauer_more_dual}
\end{eqnarray}
\begin{eqnarray}
\sum_{\gamma}
\chi^{\gamma}(b)\chi^{\gamma}(c)
=\sum_i d^{\gamma}
\chi^{\gamma}(b_i bb^{i}c),
\label{twocharacters_combine_dual}
\end{eqnarray}

\begin{eqnarray}
\sum_{i}\chi^{\gamma}_{A,\mu\nu}(b_i)
\chi^{\gamma^{\prime}}_{A^{\prime},\mu^{\prime}\nu^{\prime}}(b^{i})
=\frac{d_A}{d^{\gamma}}
\delta_{\gamma\gamma^{\prime}}
\delta_{AA^{\prime}}
\delta_{\mu\nu^{\prime}}
\delta_{\mu^{\prime}\nu},
\label{orthogonality_restricted_character_dual}
\end{eqnarray}

\begin{eqnarray}
\sum_{i}\chi^{\gamma}_{A,\mu\nu}(b_i)
\chi^{\gamma^{\prime}}_{A^{\prime},\mu^{\prime}\nu^{\prime}}(b^{i} c)
=\frac{1}{d^{\gamma}}
\delta_{\gamma\gamma^{\prime}}
\delta_{AA^{\prime}}
\chi^{\gamma}_{A,\mu\nu^{\prime}}(c)
\delta_{\mu^{\prime}\nu},
\label{orthogonality_restricted_character_Brauer_more1_dual}
\end{eqnarray}
\begin{eqnarray}
\sum_{i}\chi^{\gamma}_{A,\mu\nu}(b_i)
\chi^{\gamma^{\prime}}_{A^{\prime},\mu^{\prime}\nu^{\prime}}(cb^{i} )
=\frac{1}{d^{\gamma}}
\delta_{\gamma\gamma^{\prime}}
\delta_{AA^{\prime}}
\chi^{\gamma}_{A,\mu^{\prime}\nu}(c)
\delta_{\mu\nu^{\prime}}.
\label{orthogonality_restricted_character_Brauer_more2_dual}
\end{eqnarray}
The rewriting of orthogonality relations 
of the symmetric group using the dual basis $\sigma^i=\frac{1}{n!}\sigma_i^{-1}$ 
allows us to find  
the perfect similarity 
between the symmetric group and 
the Brauer algebra (see also \cite{Ram_thesis}).


\section{A character formula of the walled Brauer algebra}
\setcounter{equation}{0}
\label{sec:Brauer_character_formula}
In this section we will derive 
(\ref{formula_new_Brauer_character}). 
 
The character of an element $b$ in the walled Brauer algebra $B_N(m,n)$ 
is related to 
the character 
of an element of the form 
$C^{\otimes h}\otimes b_{+}\otimes b_{-}$ as 
\begin{eqnarray}
\chi^{\gamma}(b)=N^{z(b)-h}\chi^{\gamma}(C^{\otimes h}\otimes b_{+}\otimes b_{-}), 
\label{Brauer_character_general}
\end{eqnarray}
where $b_+$ is an element in $S_{m-h}$ and $b_-$ is an element in $S_{n-h}$, 
and $h$ is the number of contractions.   
The $z(b)$ denotes the number of zero cycles
\footnote{
When we write the conventional diagram of a cycle of an element $b$ 
in the walled Brauer algebra 
(for example see \cite{Ram1995,Halverson}), 
if the number of vertical edges on the left side of the wall 
minus the number of vertical edges 
on the right side of the wall 
is zero, we say that the cycle type of the cycle is zero, or 
it is a zero-cycle.
} involved in $b$. 
This formula is proved in theorem 3.1 of \cite{Ram1995}
or in 
theorem 5.13 of \cite{Halverson}. 
Note that $h$ is the number of contractions in 
$C^{\otimes h}\otimes b_{+}\otimes b_{-}$, not 
the number of contractions in $b$.  

Concretely, 
we will derive the formula for some elements in $B_N(3,3)$. Elements 
$C_{1\bar{1}}C_{2\bar{2}}$ and  
$C_{2\bar{2}}(13)(\bar{1}\bar{3})$ are already of the form 
$C^{h}\otimes b_{+}\otimes b_{-}$. On the other hand, 
$C_{1\bar{1}}C_{2\bar{2}}(23)$ and 
$C_{2\bar{2}}(123)(\bar{1}\bar{3})$ are not of the form, 
which have $z=1,0$ respectively. 
Using $C^2=NC$ and the cyclicity of the character, 
we can show 
\begin{eqnarray}
&&
\chi^{\gamma}(C_{1\bar{1}}C_{2\bar{2}}(23))
=\frac{1}{N^2}\chi^{\gamma}(C_{1\bar{1}}C_{2\bar{2}}(23)C_{2\bar{2}}C_{1\bar{1}})
=\frac{1}{N}\chi^{\gamma}(C_{1\bar{1}}C_{2\bar{2}}),
\nonumber \\
&&
\chi^{\gamma}(C_{2\bar{2}}(123)(\bar{1}\bar{3}))
=\frac{1}{N}\chi^{\gamma}(C_{2\bar{2}}(123)(\bar{1}\bar{3})C_{2\bar{2}})
=\frac{1}{N}\chi^{\gamma}(C_{2\bar{2}}(13)(\bar{1}\bar{3})),
\end{eqnarray}
which have reproduced the formula 
(\ref{Brauer_character_general}). 

For an element which is 
conjugate by some permutation in 
$S_m\times S_n$ to
the form $C^{\otimes h}\otimes b_{+}\otimes b_{-}$, 
the character of the element can be expanded 
in terms of the character in $S_{m-h}\times S_{n-h}$ as 
\begin{eqnarray}
\chi^{\gamma}(C^{\otimes h}\otimes b_{+}\otimes b_{-})
= N^{h} \sum_{\lambda \vdash (m-h)}\sum_{\pi\vdash (n-h)}
\sum_{\delta \vdash (k-h)}
g(\delta,\gamma_+;\lambda)
g(\delta,\gamma_-;\pi)
\chi_{\lambda}(b_+)
\chi_{\pi}(b_-), 
\label{character_formula2}
\end{eqnarray}
where 
$\gamma=(\gamma_+,\gamma_-)$, 
$\gamma_+\vdash (m-k)$, $\gamma_-\vdash (n-k)$. 
Note that the character vanishes if $k-h<0$. 
This formula is found in theorem 7.20 of \cite{Halverson}.
If we set $h=0$, it becomes 
a familiar formula 
\begin{eqnarray}
\chi^{\gamma}(\sigma \otimes \tau)
&=& \sum_{R \vdash m}\sum_{S\vdash n}
\sum_{\delta \vdash k}
g(\delta,\gamma_+;\lambda)
g(\delta,\gamma_-;\pi)
\chi_{R}(\sigma)
\chi_{S}(\tau)
\nonumber \\
&=&\sum_{\lambda \vdash m}\sum_{\pi\vdash n}
M^{\gamma}_{(R,S)}
\chi_{R}(\sigma)
\chi_{S}(\tau),
\end{eqnarray}
where $M^{\gamma}_{(R,S)}$ is the multiplicity of the representation $(R,S)$ 
inside the representation $\gamma$. 

In order to rewrite (\ref{character_formula2}) further, we will use   
a formula of 
Littlewood-Richardson coefficients, 
\begin{eqnarray}
g(R,S;T)=\frac{1}{n_1!n_2!}\sum_{\sigma_1\in S_{n_1}}\sum_{\sigma_2\in S_{n_2}}
\chi_R(\sigma_1^{-1})\chi_S(\sigma_2^{-1})
\chi_T(\sigma_1\circ \sigma_2),
\end{eqnarray}
where $R\vdash n_1$, $S\vdash n_2$ and $T\vdash (n_1+n_2)$. 
Substituting this into (\ref{character_formula2}), 
we obtain
\begin{eqnarray}
&&
\chi^{\gamma}(C^{\otimes h}\otimes b_{+}\otimes b_{-})
\nonumber \\
&=& N^{h} \sum_{\lambda \vdash (m-h),\pi\vdash (n-h)}
\sum_{\delta \vdash (k-h)}
\nonumber \\
&&
\times 
\frac{1}{(k-h)!(m-k)!}
\sum_{\sigma_1\in S_{k-h},\sigma_2 \in S_{m-k}}
\chi_{\delta}(\sigma_1^{-1})
\chi_{\gamma_+}(\sigma_2^{-1})
\chi_{\lambda}(\sigma_1\circ \sigma_2)
\nonumber \\
&&
\times 
\frac{1}{(k-h)!(n-k)!}
\sum_{\tau_1\in S_{k-h},\tau_2\in S_{n-k}}
\chi_{\delta}(\tau_1^{-1})
\chi_{\gamma_-}(\tau_2^{-1})
\chi_{\pi}(\tau_1\circ \tau_2)
\nonumber \\
&&
\times 
\chi_{\lambda}(b_+)
\chi_{\pi}(b_-)
\nonumber \\
&=&
N^{h} 
\frac{(m-h)!(n-h)!}{(k-h)!(m-k)!(n-k)!}
\sum_{\sigma_i,\tau_i}
\delta_{m-h}(b_+[\sigma_1\circ \sigma_2])
\delta_{n-h}(b_-[\tau_1\circ \tau_2])
\delta_{k-h}([\sigma_1^{-1}] \tau_1^{-1})
\nonumber \\
&&
\times 
\chi_{\gamma_+}(\sigma_2^{-1})
\chi_{\gamma_-}(\tau_2^{-1})
\nonumber \\
&=&N^h 
\frac{1}{(m-k)!(n-k)!}
\sum_{\sigma_2 \in S_{m-k},\tau_2\in S_{n-k}}
\Delta(b_+,b_-;\sigma_2,\tau_2)\chi_{\gamma_+}(\sigma_2^{-1})
\chi_{\gamma_-}(\tau_2^{-1}),
\label{brauer_character_symmetric_group} 
\end{eqnarray}
where we have defined 
\begin{eqnarray}
&&
\Delta(b_+,b_-;\sigma_2,\tau_2)
\nonumber \\
&=&
\frac{(m-h)!(n-h)!}{(k-h)!}
\sum_{\sigma_1,\tau_1\in S_{k-h}}
\delta_{m-h}(b_+[\sigma_1\circ \sigma_2])
\delta_{n-h}(b_-[\tau_1\circ \tau_2])
\delta_{k-h}([\sigma_1^{-1}] \tau_1^{-1})
\nonumber \\
&=&
\frac{1}{((k-h)!)^2}
\sum_{\sigma_1,\tau_1\in S_{k-h}}
tr^{(r)}_{m-h}(b_+[\sigma_1\circ \sigma_2])
tr^{(r)}_{n-h}(b_-[\tau_1\circ \tau_2])
tr^{(r)}_{k-h}([\sigma_1^{-1}] \tau_1^{-1}).
\end{eqnarray}
Note again that $b_+\in S_{m-h}$ and  
$b_- \in S_{n-h}$, while 
$\sigma_2 \in S_{m-k}$ and  
$\tau_2 \in S_{n-k}$. 
$\delta_{l}(\sigma)$ is the delta function defined over the group algebra of $S_l$. 
To obtain the second equality in 
(\ref{brauer_character_symmetric_group}) we have used the formula 
(\ref{character_sum_R_formula}), 
and 
$[\sigma]$ is defined in (\ref{conjugacy_class_symbol}). 
We have introduced the trace of the regular representation of $S_n$ as 
\begin{eqnarray}
tr^{(r)}_n(\sigma)=n!\delta_n(\sigma) \quad (\sigma\in S_n). 
\end{eqnarray}
If we use the two-point function of the topological field theory given 
in section \ref{sec:TFT_review}, we have
\begin{eqnarray}
\Delta(b_+,b_-;\sigma_2,\tau_2)
=\frac{1}{(k-h)!^2}
\sum_{\sigma_1,\tau_1\in S_{k-h}}
\eta^{S_{m-h}}(b_+,\sigma_1\circ \sigma_2)
\eta^{S_{n-h}}(b_-,\tau_1\circ \tau_2)
\eta^{S_{k-h}}(\sigma_1^{-1}, \tau_1^{-1}). 
\label{Delta_TFT_cylinders}
\end{eqnarray}
It is convenient to keep in our mind that 
$\sigma^{-1}\in S_n$ always appears with $1/n!$. Using the dual basis 
$\sigma^i=\frac{1}{n!}\sigma_i^{-1}$, we get tidy expressions without 
factorials; 
\begin{eqnarray}
\chi^{\gamma}(C^{\otimes h}\otimes b_{+}\otimes b_{-})
=N^h 
\sum_{i\in S_{m-k},j\in S_{n-k}}
\Delta(b_+,b_-;\sigma_i,\tau_j)\chi_{\gamma_+}(\sigma^i)
\chi_{\gamma_-}(\tau^j),
\end{eqnarray}
and
\begin{eqnarray}
\Delta(b_+,b_-;\sigma_i,\tau_j)
=
\sum_{k,l\in S_{k-h}}
\eta^{S_{m-h}}(b_+,\sigma_k\circ \sigma_i)
\eta^{S_{n-h}}(b_-,\tau_l\circ \tau_j)
\eta^{S_{k-h}}(\sigma^k, \tau^l). 
\end{eqnarray}
These look like pleasant but having too many indices, which might confuse 
the readers, so we do not use these 
expressions in the main text. 


\end{document}